\documentclass[12pt]{article}
\usepackage{a4}
\usepackage{latexsym} 
\usepackage{amssymb}  
\usepackage{amsmath} 
\usepackage{graphicx}
\usepackage{epsfig,wrapfig}
\usepackage[ usenames]{color}
\usepackage[rflt]{floatflt} 
\begin{document}

\newcommand{\talk}[3]
{\noindent{#1}\\ \mbox{}\ \ \ {\it #2} \dotfill {\pageref{#3}}\\[1.8mm]}
\newcommand{\stalk}[3]
{{#1} & {\it #2} & {\pageref{#3}}\\}
\newcommand{\snotalk}[3]
{{#1} & {\it #2} & {{#3}n.r.}\\}
\newcommand{\notalk}[3]
{\noindent{#1}\\ \mbox{}\ \ \ {\it #2} \hfill {{#3}n.r.}\\[-4mm]}
\newcounter{zyxabstract}     
\newcounter{zyxrefers}        

\newcommand{\newabstract}
{\newpage\stepcounter{zyxabstract}\setcounter{equation}{0}
\setcounter{footnote}{0}\setcounter{figure}{0}\setcounter{table}{0}}

\newcommand{\rlabel}[1]{\label{zyx\arabic{zyxabstract}#1}}
\newcommand{\rref}[1]{\ref{zyx\arabic{zyxabstract}#1}}

\renewenvironment{thebibliography}[1] 
{\section*{References}\setcounter{zyxrefers}{0}
\begin{list}{ [\arabic{zyxrefers}]}{\usecounter{zyxrefers}}}
{\end{list}}
\newenvironment{thebibliographynotitle}[1] 
{\setcounter{zyxrefers}{0}
\begin{list}{ [\arabic{zyxrefers}]}
{\usecounter{zyxrefers}\setlength{\itemsep}{-2mm}}}
{\end{list}}

\renewcommand{\bibitem}[1]{\item\rlabel{y#1}}
\renewcommand{\cite}[1]{[\rref{y#1}]}      
\newcommand{\citetwo}[2]{[\rref{y#1},\rref{y#2}]}
\newcommand{\citethree}[3]{[\rref{y#1},\rref{y#2},\rref{y#3}]}
\newcommand{\citefour}[4]{[\rref{y#1},\rref{y#2},\rref{y#3},\rref{y#4}]}
\newcommand{\citefive}[5]
{[\rref{y#1},\rref{y#2},\rref{y#3},\rref{y#4},\rref{y#5}]}
\newcommand{\citesix}[6]
{[\rref{y#1},\rref{y#2},\rref{y#3},\rref{y#4},\rref{y#5},\rref{y#6}]}

\begin{titlepage}

\begin{flushleft}
\includegraphics[height=1.5cm]{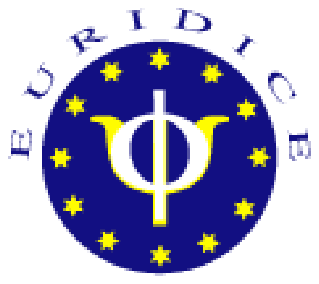} 
\end{flushleft}

\vspace*{-1.75cm}
\begin{flushright}
\small{\tt ECT*-04-01, HISKP-TH-04/02}
\end{flushright}

\vspace*{1.cm}

\begin{center}
{\large \bf HadAtom03}\\[0.5cm]
{Workshop on Hadronic Atoms,}\\
{ECT*, Strada delle Tabarelle 286, I-38050, Villazzano (Trento), Italy}\\
{October 13-17, 2003}
\\[1cm]
{\em edited by}\\[1cm]
{\bf J. Gasser$^1$, A. Rusetsky$^{2,3}$, J. Schacher$^4$}\\[0.3cm]
{\em $^1$Institute for Theoretical Physics, 
University of Bern, Sidlerstrasse 5, 3012 Bern, Switzerland}\\
{\em $^2$Universit\"{a}t Bonn, Helmholtz-Institut f\"{u}r
Strahlen- und Kernphysik (Theorie), Nu{\ss}allee 14-16, D-53115 Bonn, Germany}\\
{\em $^3$On leave of absence from: 
HEPI, Tbilisi State University, University st. 9, 380086 Tbilisi, Georgia}\\
{\em $^4$Laboratory for High-Energy Physics, University of Bern, Sidlerstrasse 5, 3012 Bern, Switzerland}

\begin{center}
\includegraphics[height=5.cm]{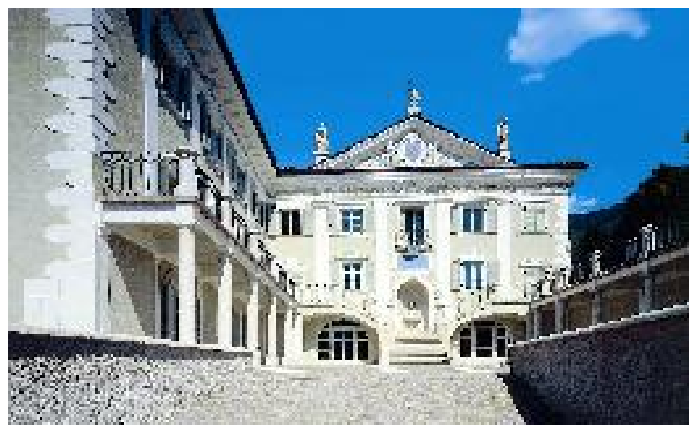} 
\end{center}

{\large ABSTRACT}
\end{center}
These are the proceedings of the workshop ``HadAtom03'', held at
the European Centre for Theoretical Nuclear Physics and Related
Studies (ECT*), Trento (Italy), October 13 - 17, 2003.
The main topics discussed at the  workshop were the physics of 
hadronic atoms and in 
this context recent results in experiment and theory.
Included here are the list of participants, the scientific program and
a short contribution from each speaker.

\end{titlepage}

\setcounter{page}{2}

\newabstract

\section{Introduction}

\vspace{1cm}

The HadAtom03 workshop was held on October 13-17, 2003 at the European Centre
for Theoretical Nuclear Physics and Related Topics (ECT*), Trento (Italy).
This was already the fifth regular workshop of the HadAtom series, which 
originally have been inspired by latest experimental and
theoretical progress achieved in the investigation of the bound states of
strongly interacting particles -- hadronic atoms. The previous workshops
were held
in Dubna (1998), Bern (1999,2001), and at CERN (2002). At this time, the topics
of the workshop included:

\begin{itemize}
\itemsep -1.mm
\item
Hadronic atoms, in particular their

\begin{itemize}
\itemsep -1.5mm
 \item               Production
 \item               Interaction with matter
 \item               Energy levels
 \item               Decays
\end{itemize}

\item
Meson-meson and meson-baryon scattering

\item
Experiments

\begin{itemize}
\itemsep -1.5mm
\item                DIRAC at CERN
\item                DEAR at DAFNE
\item                Pionic Hydrogen Collaboration at PSI
\item                Experiment on deeply bound pionic atoms at GSI
\item                Others
\end{itemize}

\item
Lattice calculations and effective theory of QCD

\item
Chiral Perturbation Theory and nuclear many-body systems

\begin{itemize}
\itemsep -1.5mm
\item                Nuclear Matter
\item                Finite systems
\item                $\pi$-nucleus, $K$-nucleus bound states: spectrum and decays
\end{itemize}
\end{itemize}

More than 40 physicists took part in the workshop, and 35 talks were 
presented.  For the first time during the workshop we allocated 
free slots for informal discussions,
particularly useful for not being restricted with time limits and
a detailed program.
The topics and the coordinators of the discussions were announced in advance.
In total, three discussions took place during the workshop:

\begin{tabbing}
Topicxxxxxxxxxxxxxxxxxxxxxxxxxxxxxxxxxxxxxxxxxxxxxx\=Coordinationxxxxxxxxx\kill
{\em Topic} \> {\em Coordination} \\[2mm]
$\pi K$ interaction and three-flavour ChPT\>Stern\\[2mm]
The single arm results from DIRAC: \> \\
pion propagation in the medium \> Nemenov and Wirzba\\[2mm]
Spectrum and decays of the $KN$ atom\> Petrascu and Rusetsky
\end{tabbing}

As after the previous workshops \citefour{98}{99}{01}{02}, we publish a collection
of abstracts of the presentations, containing 
relevant references. In addition, we display 
the list of the participants with their e-mail addresses.

\bigskip\bigskip

{\em Acknowledgments.}

We wish to thank all participants for travelling to Trento and for making
``HadAtom03'' an exciting and lively meeting. We want to thank our secretary
Ines Campo for the excellent performance, and the staff of the ECT*:
Corrado Carlin, Cristina Costa, Barbara Curro-Dossi, Tiziana Ingrassia, 
Mauro Meneghin, Donatella Rossetti, Rachel Weatherhead and 
Gianmaria Ziglio for their help. Last but not least, we thank our colleagues
from the program panel: Leonid Afanasyev, Valery Lyubovitskij, Leonid Nemenov,
Hagop Sazdjian and Dirk Trautmann, for their invaluable contribution in 
structuring the meeting. 

The ``HadAtom03'' workshop was partially supported by the ECT*,
and by RTN, BBW-Contract No.01.0357
and EC-Contract  HPRN--CT2002--00311 (EURIDICE).
It was a topical workshop of the EURIDICE collaboration.

\bigskip

\noindent Bern/Bonn, January 26, 2004

\bigskip\bigskip\bigskip

\hspace*{10.cm} J\"urg Gasser

\hspace*{10.cm} Akaki Rusetsky

\hspace*{10.cm} J\"urg Schacher

\newpage
\section{List of Particpants}

{\sf
\begin{tabbing}
Noxx\=A very long namexxxxx\=a very long institutexxxxxx\=email\kill
1.\>L. Afanasyev  \>(Dubna) \> leonid.afanasev@cern.ch\\                       
2.\>P. B\"uttiker \> (Bonn)  \> buettike@itkp.uni-bonn.de\\                         
3.\>V. Brekhovskikh  \>(Dubna) \>valeri.brekhovskikh@cern.ch \\                      
4.\>M. Cargnelli \> (Vienna) \> Michael.Cargnelli@oeaw.ac.at   \\                     
5.\>A. Dax \> (PSI) \> andreas.dax@psi.ch   \\                              
6.\>D. Drijard \> (CERN) \> daniel.drijard@cern.ch \\                           
7.\>S. D\"urr \> (Zeuthen) \>stephan.duerr@desy.de \\                           
8.\>T. Ericson \> (CERN) \> torleif.ericson@cern.ch \\                           
9.\>E. Friedman \> (Jerusalem) \>elifried@vms.huji.ac.il \\                      
10.\>J. Gasser \> (Bern) \> gasser@itp.unibe.ch   \\                          
11.\>L. Girlanda \> (Trento) \> girlanda@ect.it \\                        
12.\>D. Gotta \> (J\"ulich) \> d.gotta@fz-juelich.de \\                          
13.\>K. Hencken \> (Basel) \> k.hencken@unibas.ch\\                           
14.\>S. Hirenzaki  \>(Nara Woman's College) \>zaki@phys.nara-wu.ac.jp \\          
15.\>A. Ivanov \> (Vienna/St.Petersburg) \>ivanov@kph.tuwien.ac.at \\             
16.\>T. Jensen \> (Paris) \> thomas.jensen@spectro.jussieu.fr\\                            
17.\>D. Jido \> (Trento) \> jido@ect.it \\                            
18.\>J. Juge \> (Dublin) \> juge@itp.unibe.ch \\                            
19.\>S. Krewald  \>(J\"ulich) \> s.krewald@fz-juelich.de \\ 
20.\>E. Kolomeitsev \> (Nordita) \> e.kolomeitsev@gsi.de \\                       
21.\>R. Lemmer \> (Johannesburg) \> LemmerR@physics.wits.ac.za\\                     
22.\>E. Lipartia \> (Lund) \> lipartia@thep.lu.se  \\                         
23.\>L. Ludhova \> (Fribourg) \> livia.ludhova@unifr.ch\\
24.\>V. Lyubovitskij \> (T\"ubingen) \> valeri.lyubovitskij@uni-tuebingen.de\\                  
25.\>H. Nagahiro \> (Nara Woman's College) \> h.nagahiro@cc.nara-wu.ac.jp  \\         
26.\>L. Nemenov \> (CERN/Dubna) \>  leonid.nemenov@cern.ch  \\                   
27.\>G. Pancheri  \>(LNF-INFN) \> giulia.pancheri@lnf.infn.it \\                      
28.\>C. Petrascu \> (LNF-INFN) \>catalina@lnf.infn.it  \\                      
29.\>U. Raha  \>(Bonn) \>  udit@itkp.uni-bonn.de  \\                            
30.\>A. Rusetsky \> (Bonn/Tbilisi) \> rusetsky@itkp.uni-bonn.de \\                  
31.\>M. Sainio \> (Helsinki) \>  mikko.sainio@helsinki.fi  \\                      
32.\>H. Sazdjian \> (Paris) \> sazdjian@ipno.in2p3.fr \\                         
33.\>J. Schacher \> (Bern) \> schacher@lhep.unibe.ch\\                           
34.\>J. Schweizer \> (Bern) \> schweize@itp.unibe.ch \\                          
35.\>L. Simons \> (PSI) \> leopold.simons@psi.ch  \\                            
36.\>J. Stern \> (Paris) \> stern@ipno.in2p3.fr \\                            
37.\>K. Suzuki \> (M\"unchen) \> ken.suzuki@ph.tum.de\\                         
38.\>A. Tarasov  \>(Dubna) \>avt@mpimail.mpi-hd.mpg.de \\                           
39.\>D. Trautmann \> (Basel) \>dirk.trautmann@unibas.ch \\                         
40.\>O. Voskresenskaya \> (Dubna) \> voskr@mpimail.mpi-hd.mpg.de\\                    
41.\>W. Weise \> (Trento/Munich)  \> weise@ect.it\\                     
42.\>A. Wirzba \> (Bonn)  \> wirzba@itkp.uni-bonn.de \\                           
43.\>P. Zemp \> (Bern) \> zemp@itp.unibe.ch   
\end{tabbing}

}

\newpage

\section{Contributions}

\vskip.5cm

\noindent\mbox{}\hfill{\bf Page}

\talk{{\bf L. Nemenov}}{Atoms consisting of $\pi^+\pi^-$ and $\pi K$ mesons}
{abs:Nemenov}
\talk{{\bf V. Brekhovskikh}}{Observation of $\pi^+\pi^-$ atoms at DIRAC and its 
lifetime estimation}{abs:Brekhovskikh}
\talk{{\bf M.E. Sainio}}{Pion-nucleon scattering}{abs:Sainio}
\talk{{\bf A. Tarasov} and L. Afanasyev}
{Moliere multiple-scattering theory revisited}{abs:Tarasov1}
\talk{M. Schumann, T. Heim, {\bf K. Hencken}, D. Trautmann, and 
G. Baur}{Interaction of Pionium with atoms at relativistic energies and
propagation of Pionium through matter}{abs:Hencken}
\talk{{\bf L. Afanasyev} and L. Nemenov}{Production of the long-lived 
excited states of $\pi^+\pi^-$ atoms in view to perform an experiment on 
their observation and measurement the Lamb shift}{abs:Afanasyev}
 \talk{{\bf A. Tarasov} and O. Voskresenskaya}{The quantum-mechanical 
treatment of pionium internal dynamics in matter}{abs:Tarasov2}
\talk{{\bf A. N. Ivanov}, M. Cargnelli, M. Faber,
A. Hirtl, J. Marton, N. I. Troitskaya, and J. Zmeskal}{On radiative 
decay channels of pionic and kaonic hydrogen}{abs:Ivanov}
\talk{{\bf P. Zemp}}{Deser-type formula for pionic hydrogen}{abs:Zemp}
\talk{{\bf T.E.O.~Ericson},  B.~Loiseau, and S.~Wycech}{Electromagnetic 
corrections to  scattering lengths from hydrogenic atoms applied to the 
$\pi ^-$p system}{abs:Ericson}
\talk{{\bf D. Gotta}}{Ground-State Shift in Pionic hydrogen}{abs:Gotta}
\talk{{\bf T.S. Jensen}}{Cascade model predictions: $K^-p$, $K^-d$, and $\pi^-p$}{abs:Jensen}
\talk{{\bf L.M. Simons}}{Experimental determination of the strong interaction 
ground state width in pionic\\ hydrogen}{abs:Simons}
\talk{{\bf H. Sazdjian}}{Calculating $\pi K$ atom properties
in the constraint theory approach}{abs:Sazdjian}
\talk{{\bf J. Gasser}}{ On the precision of the theoretical predictions for
  $\pi\pi$ scattering}{abs:Gasser}
\talk{{\bf J. Schweizer}}{Energy shift and decay width of the $\pi K$ atom}{abs:Schweizer}
\talk{{\bf P. B\"uttiker}, S. Descotes--Genon, and B. Moussallam}
{Roy--Steiner equations for $\pi K$ scattering}{abs:Buettiker}
\talk{S.~Descotes-Genon, N.~H.~Fuchs, {\bf L. Girlanda}, and J.~Stern}
{The impact of $\pi \pi$ scattering data on SU(3) chiral dynamics}{abs:Girlanda}
\talk{{\bf S. Hirenzaki}}{Formation of Meson-Nucleus Systems}{abs:Hirenzaki}
\talk{{\bf H. Nagahiro}, D.~Jido, and S.~Hirenzaki}
{$\eta$-Nucleus interactions and in-medium properties of 
$N^*(1535)$ in chiral models}{abs:Nagahiro}
\talk{{\bf K. Suzuki}}{Precise measurement of deeply bound pionic 1s states of Sn nuclei}{abs:Suzuki}
\talk{{\bf M. Cargnelli} {\em et al.}}{DEAR - Kaonic Hydrogen: First Results}
{abs:Cargnelli}
\talk{{\bf E.E.~Kolomeitsev}, N.~Kaiser, and W.~Weise}
{Chiral dynamics and  pionic states of Pb and Sn isotopes}{abs:Kolomeitsev}
\talk{{\bf E. Friedman} and A. Gal}{Chiral restoration from pionic atoms}{abs:Friedman}
\talk{L. Girlanda, {\bf A. Rusetsky}, and W. Weise}
{Introduction to the ChPT for Heavy Nuclei}{abs:Rusetsky}
\talk{{\bf A. Wirzba}}{ChPT in the nuclear medium - the generating functional 
approach}{abs:Wirzba}
\talk{{\bf C. Curceanu (Petrascu)}}{Future precision measurements on kaonic
 hydrogen and kaonic deuterium with\\ SIDDHARTA}{abs:Petrascu}
\talk{{\bf S. Krewald}, R.H. Lemmer, and F.P.  Sassen}{Lifetime of Kaonium}{abs:Krewald}
\talk{{\bf D. Jido},  J.A. Oller, E. Oset, A. Ramos, and U.-G. Mei\ss{}ner}
{Chiral dynamics of the two $\Lambda(1405)$ states}{abs:Jido}
\talk{{\bf V.\ E.\ Lyubovitskij}, Amand \ Faessler, Th.\ Gutsche, 
K. \ Pumsa-ard, P. \ Wang}{Chiral dynamics of baryons as bound states of  
constituent quarks}{abs:Lyubovitskij}
\talk{{\bf A. Dax} {\em et al.}}{Towards the most precise test of bound state QED}{abs:Dax}
\talk{{\bf K.J.~Juge}}{Lattice QCD calculations of the I=2 pi-pi scattering length}{abs:Juge}
\talk{{\bf S. D\"urr}}{Towards a lattice determination of QCD low-energy constants}{abs:Duerr}
\talk{M. Procura, T. Hemmert, and {\bf W. Weise}}
{Nucleon mass, sigma term and lattice QCD: chiral extrapolations}{abs:Weise}
\hspace*{1.cm}


\newabstract 
\label{abs:Nemenov}

\begin{center}
{\large\bf Atoms consisting of $\pi^+\pi^-$ and $\pi K$ mesons}\\[0.5cm]
L.Nemenov$^1$\\[0.3cm]
$^1$JINR, Dubna\\[0.3cm]
\end{center}

Using experience obtained in the DIRAC experiment at CERN \cite{[1]} the 
new experimentis proposed for the PS CERN and J-PARC in Japan  \cite{[2]} 
to perform a crucial check of the precise predictions of the low energy
QCD. The DIRAC Collaboration plans to upgrade the setup: 1) improving the
detector shielding, 2) adding aerogel counters in combination with existing
Cerenkov counters to identify K-mesons and protons and 3) modernising
part of the electronics.

The proposed experiment aims to measure the lifetime of $\pi^+\pi^-$-atom
($A_{2\pi}$) with precision better then 6\% and the determination of the
difference S-wave pion-pion scattering lengths $|a_0~-~a_2|$ at the level
of 3\%. Simultaneously with investigation of $A_{2\pi}$ with the same setup 
the observation of atoms consisting of $\pi K$-mesons ($A_{\pi K}$) and their
lifetime measurement with precision of $\sim$20\% and the evaluation of the
difference S-wave $\pi K$ scattering lengths $|a_{1/2}~-~a_{3/2}|$ with
precision $\sim$10\% is planned.

The observation of the long-lived (metastable) states of $A_{2\pi}$ is also 
planned with the same setup. This will give the possibility to measure the  
difference $\Delta E_n$ between energy of nS and nP states and to evaluate
in a model independent way the value  $2a_0 +a_2$. The measurements of the
$A_{2\pi}$~lifetime and $\Delta E_n$ allow to obtain in a model independent
way values for $a_0$ and $a_2$, separately.

Low energy QCD \cite{[3]} predicts for the pion-pion scattering lengths with 
an accuracy about 2\% ~\citetwo{[4]}{[5]} and about 10\% of the pi-K scattering
lengths \citetwo{[6]}{[7]}. The pion-pion and pion-kaon scattering lengths have 
never been verified by the experimental data with the same accuracy as the 
theoretical prediction.

These theoretical results have been obtained assuming strong
condensation of the quark-antiquark pairs in the vacuum. On this reason
the proposed experiment will be a crucial check of the low energy QCD
predictions and our understanding of the nature of QCD vacuum \cite{[8]}.

\newabstract 
\label{abs:Brekhovskikh}

\begin{center}
{\large\bf Observation of $\pi^+\pi^-$ atoms at DIRAC and its lifetime estimation}\\[0.5cm]
{\bf V. Brekhovskikh} on behalf of the DIRAC collaboration\\[0.3cm]
Institute for High Energy Physics, Pobedy 1, 142284 Protvino, Russia\\[0.1cm]
\end{center}

The Chiral Perturbation Theory (ChPT) relies on the 2-loop corrections
and Roy equations to give an accurate prediction of the difference
between the S-wave $\pi\pi$~scattering isospin 0 and 2 lengths, $a_0$ and
$a_2$. $|a_0-a_2|$ is predicted with an accuracy of 1.5\%: $0.265 \pm
0.004$ \cite{[1]}. The dominant mode of $\pi^+\pi^-$~atom ($A_{2\pi}$) decay is 
$A_{2\pi}\rightarrow \pi^0\pi^0$, whereas the partial width of the second decay 
mode $A_{2\pi}\rightarrow \gamma\gamma$ is 0.4\%. The relationship 
between the decay width $\Gamma_{2\pi^0}$ and the $\pi\pi$ scattering 
lengths is $\Gamma_{2\pi^0} = C\cdot|a_0-a_2|^2$ ~\citetwo{[2]}{[3]}. 
The next-to-leading order correction for the isospin breaking gives: 
$\Gamma^{NLO}_{2\pi^0} = \Gamma_{2\pi^0}(1+\delta_\Gamma)$, 
with $\delta_\Gamma = (5.8 \pm 1.2)\%$ \cite{[1]}.  This yields the value 
of the $A_{2\pi}$~lifetime $\tau = (2.9 \pm 0.1) \times 10^{-15}$ s. 
A measurement of the latter with an accuracy of 10\% allows one to
determine the corresponding $\pi\pi$ scattering length difference with
a 5\% precision.

The following method is used for the lifetime measurement by the DIRAC
experiment. A $\pi^+\pi^-$~pair are generated from the short-lived sources
as a result of the proton-target interaction.  Final state
$\pi^+\pi^-$~Coulomb interaction may lead to the creation of an atom $A_{2\pi}$~ 
\citetwo{[5]}{[6]}. Once created, the $A_{2\pi}$~propagates through the 
target and may break up due to interaction with target matter,
producing so-called "atomic pairs". These pairs are characterized by small
relative momenta $Q<3$ MeV/c in the pair c. m. system. A further process
competing with the breakup is the $A_{2\pi}$~annihilation into $\pi^0\pi^0$.
Additional $\pi^+\pi^-$~pairs originating from the short lived sources can form
free "Coulomb pairs" \cite{[6]}, e.g. pion pairs in a free or continuous 
state. The number "Coulomb pairs" is related to the number of produced 
$A_{2\pi}$'s. With the definition of the breakup probability $P_{br}$ as 
the ratio of the number of dissociated atoms $n_A$ to the number of 
produced atoms, a lifetime vs. breakup probability relationship may be 
established by solving a set of differential transport equations. 
Then $P_{br}$ can be derived with an accuracy of better than 1\% \cite{[7]}.

The DIRAC spectrometer was designed specifically for detecting
$\pi^+\pi^-$~pairs with small relative momenta \cite{[8]}. The setup is located
in the T8 beam area of the CERN PS accelerator. Three detectors 
are installed upstream of the magnet, and six detectors in 
two identical arms downstream of the magnet. The setup resolution for 
the relative c. m. momentum $Q$ for $\pi^+\pi^-$~pairs is better than 1 MeV/c.

The method of extracting "atomic pairs" is explained in more detail in 
\cite{[9]}. The accuracy of this method is defined not only by the 
statistical errors, but also the systematic one.  There are two main 
sources of the latter. The first one is the accuracy of the description 
of multiple scattering in the target and detectors and the second 
one is the scintillating  fiber detector response for two close tracks. 
To this end we have dedicated portions of our runs in 2002-2003 to the 
investigation of the systematic errors. During 2002-2003 runs, 
two Ni targets, single and multilayer, were also used. Single and 
multilayer target event distributions are identical in all respects 
but one: the multilayer target yields a lower number of dissociated 
pairs due to the annihilations in the interlayer gaps. 
This allows one immediately obtain the atomic pair signal from the 
difference between the single and multilayer target distributions.

It is possible to decrease the systematic error by a factor of 2 if only 
the longitudinal momentum component $Q_{L}$ is taken into consideration 
\cite{[9]}. An alternative approach for the signal extraction is 
to make use of the downstream detectors only. This method allows us 
to get a more realistic number of "atomic pairs" due to absence 
of all criteria on hit occupancy and inefficiency in upstream 
detectors. In Fig. 1 we present the $Q_{L}$ distributions of "atomic 
pairs" with and without the information from the upstream detectors.

\begin{figure}[h]
\begin{minipage}[b]{0.47\textwidth}
\begin{center}
\includegraphics[width=6.cm]{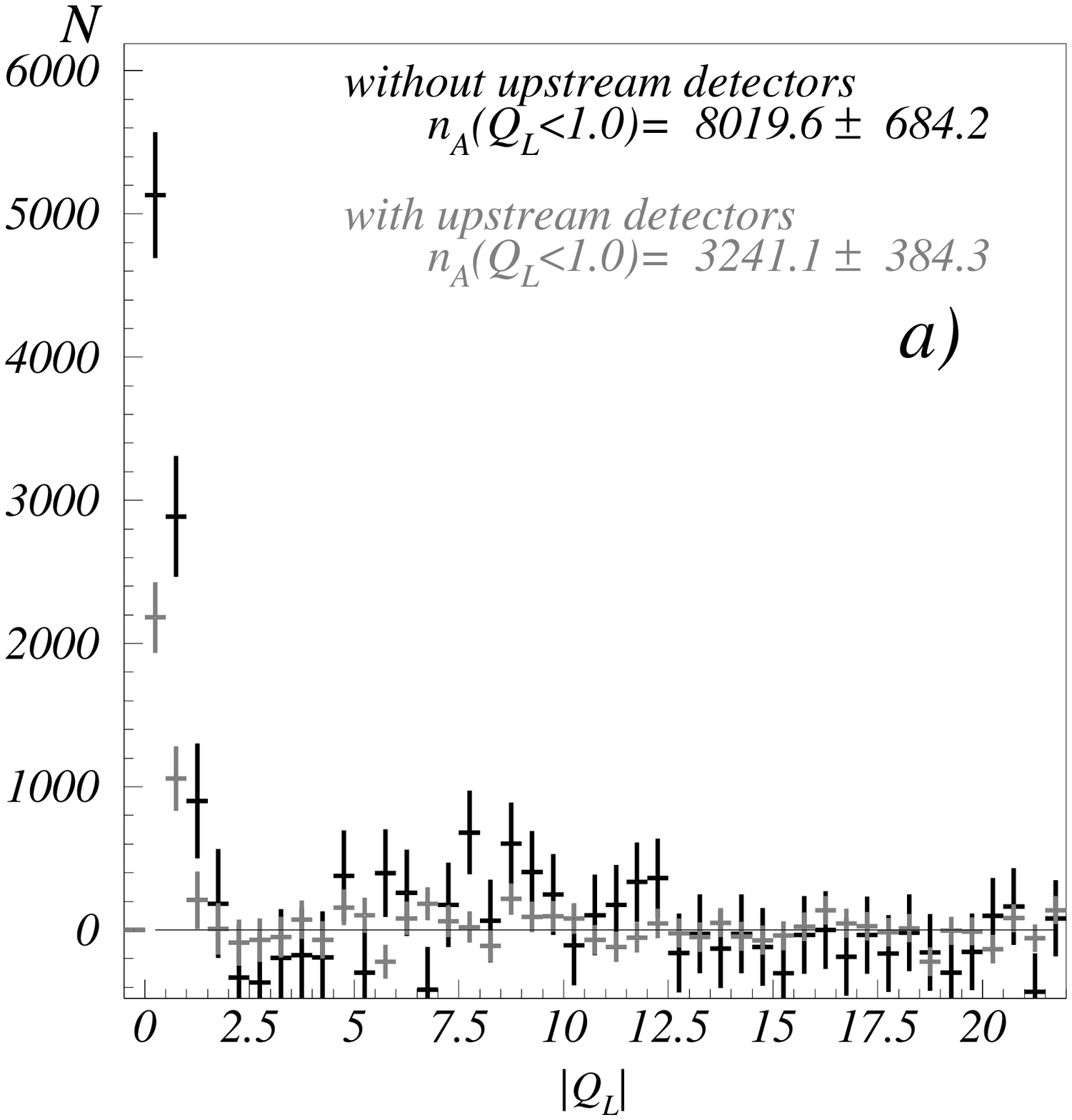}
\end{center}
\end{minipage}\hfill%
\begin{minipage}[b]{0.47\textwidth}
\begin{center}
\includegraphics[width=6.cm]{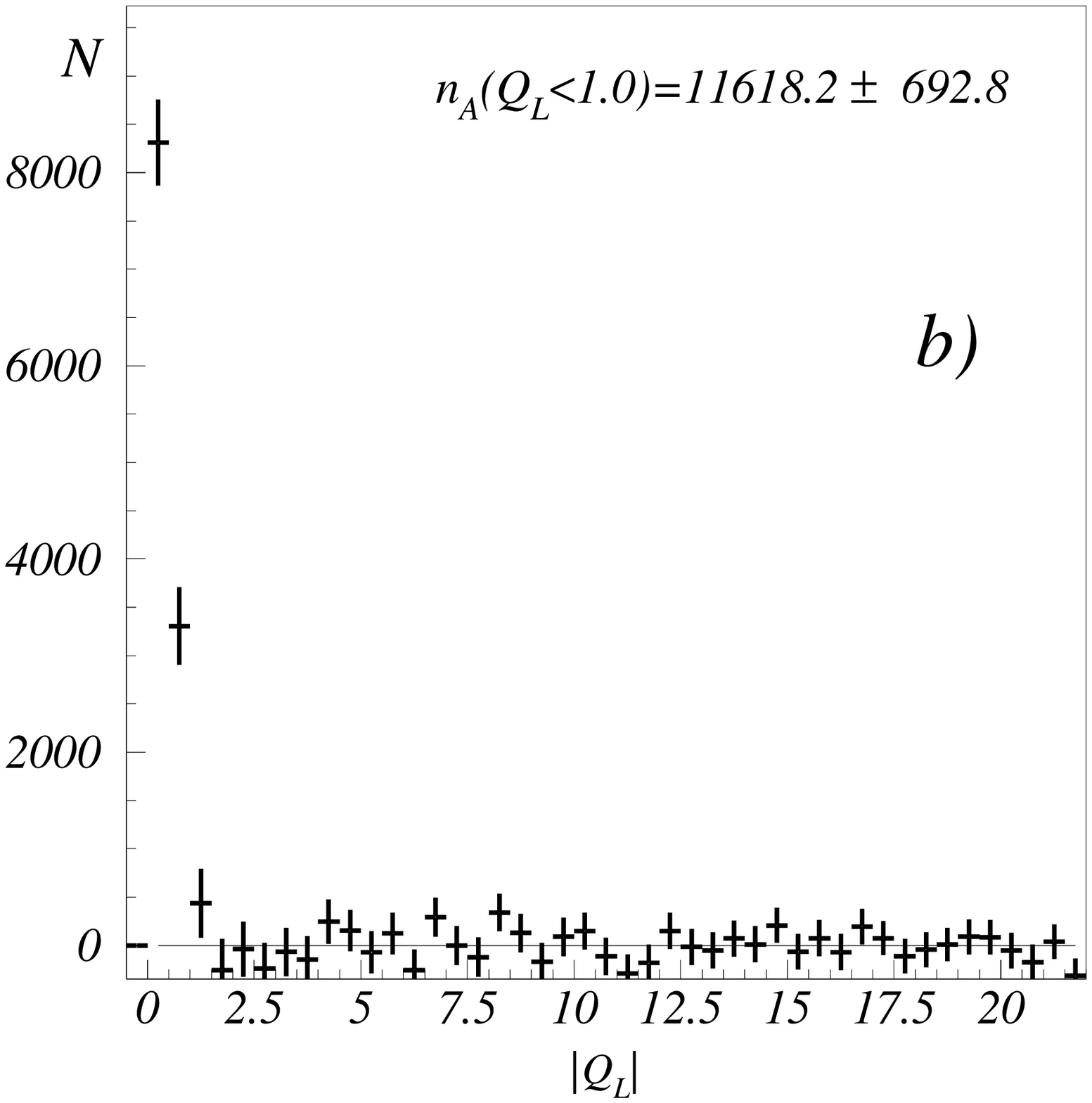}
\end{center}
\end{minipage}
\caption{"Atomic pair" signals: a) for the combined Ni 2000 and Ti
2000-2001 statistics ($\sim$30\% of all statistics) with and without 
upstream detectors and b) for the combined 
Ni 2001-2003 statistics ($\sim$70\% of all statistics) with upstream 
detectors.}
\label{fig1}
\end{figure}

The total number of "atomic pairs" collected during 2000-2003 runs is 
more than 20000 in the case of identification using downstream detectors 
only.

The data analysis procedure has not been finalized. Taking into
account the best understood part of the Ni 2001 data statistics with 
about 5200 "atomic pairs", a preliminary estimate for the lifetime yields 
the value: $\tau =[3.1^{+0.9}_{-0.7}(stat) \pm 1 (syst)] 
\times 10^{-15}$ s. The analysis of the data collected in 2000 and 2003 
allows us to reduce the overall error down to 12-14\%.

\newabstract 
\label{abs:Sainio}

\begin{center}
{\large\bf Pion-nucleon scattering}\\[0.5cm]
M.E. Sainio\\[0.3cm]
Helsinki Institute of Physics and Department of Physical Sciences, 
University of Helsinki, Finland
\end{center}

The meson factories at Los Alamos (LAMPF), Vancouver (TRIUMF) and Villigen
(PSI) have produced a lot of pion-nucleon scattering data at low and medium 
energies. Most of these data were not yet available at the time of the Karlsruhe-Helsinki
partial wave analysis \cite{KH80:1980}, but are included in the data dase of the
SAID analysis \cite{said}. The new data are expected to change the low-energy
pion-nucleon partial wave amplitudes. It is our aim to perform a pion-nucleon partial
wave analysis in the spirit of the Karlsruhe-Helsinki approach by making use of the
expansion method of Pietarinen and to include the data of the meson factory era.

Examples of recent data sets at low energy are the data of Pavan et al. \cite{pavan:2001}
on differential cross sections, Hofman et al. \cite{hofman:2003} on polarization,
Alekseev et al. \cite{alekseev:2001} on spin rotation parameters and Kriss et al. \cite{kriss:1999}
on integrated cross sections. In addition, there are precise measurements of the level shift and
width of the pionic hydrogen. Also, the CHAOS collaboration at TRIUMF has measured elastic 
scattering at the Coulomb-nuclear interference region providing information for the forward
extrapolation \cite{meier:2002}.

The electromagnetic corrections will be treated in the manner of Tromborg et al. \cite{Tromborg:1977}
eventhough the approach is known to have shortcomings \cite{Fettes:2001}.

Currently we are still testing the analysis programme with theoretical ``data'' from the
KA85 analysis \cite{koch:1986}.

\newabstract 
\label{abs:Tarasov1}

\begin{center}
  {\large\bf Moliere multiple-scattering theory revisited}\\[0.5cm]
{\bf A. Tarasov} and L. Afanasyev\\[0.3cm]
Joint Institute for Nuclear Research, 141980 Dubna,Moscow Region, Russia
\end{center}

We have derived the rigorous relation between the values of the
screening parameters (angle) $\theta_a$ of the Moliere multiple
scattering theory \cite{moliere}, calculated in the Born and WKB
approximations, instead of the approximate one obtained in the
original paper by Moliere.

Our result reads:

$$
\ln (\theta_a^{\mathrm{WKB}})- \ln (\theta_a^{\mathrm{Born}}) = 
\mathrm{Re} \left[\psi(1+i\alpha) -\psi(1)\right]
$$
$$
\alpha=\frac{Z}{137\beta} \qquad \psi(x)=\frac{d}{dx} \ln \Gamma(x) \,.
$$

Here $Z$ is the atomic number of the target atom and $\beta$ is the
charged particle velocity.

\newabstract 
\label{abs:Hencken}

\begin{center}
{\large\bf Interaction of Pionium with atoms at relativistic energies and
propagation of Pionium through matter}\\[0.5cm]
M. Schumann$^1$, T. Heim $^1$, {\bf K. Hencken}$^1$, D. Trautmann$^1$ and 
G. Baur$^2$\\[0.3cm]
$^1$Inst.\ f\"ur Physik, University of Basel, Klingelbergstr.\ 82 ,
4056 Basel, Switzerland\\[0.1cm]
$^2$IKP (Theorie), Forschungszentrum J\"ulich, 52425 J\"ulich, Germany
\end{center}

In order to be able to measure the pionium ($A_{2\pi}$) lifetime with an
accuracy of 10\%, the experiment DIRAC at CERN needs as theoretical input
the electromagnetic excitation and ionization cross sections of this exotic 
atom with 1\% accuracy. This is due to the fact, that DIRAC determines the 
lifetime indirectly by measuring the number of pionium atoms breaking up 
(in contrast to being annihilated to $\pi^0\pi^0$) and the number of produced 
atoms. From this breakup ratio, together with its theoretical
calculation  as a function of the lifetime of pionium, the lifetime is
estimated, see Fig.~\ref{figmixed}(A) below.

\begin{figure}[ht]
\begin{minipage}[b]{0.48\hsize}
\leavevmode
\centerline{\includegraphics[width=5.2cm]{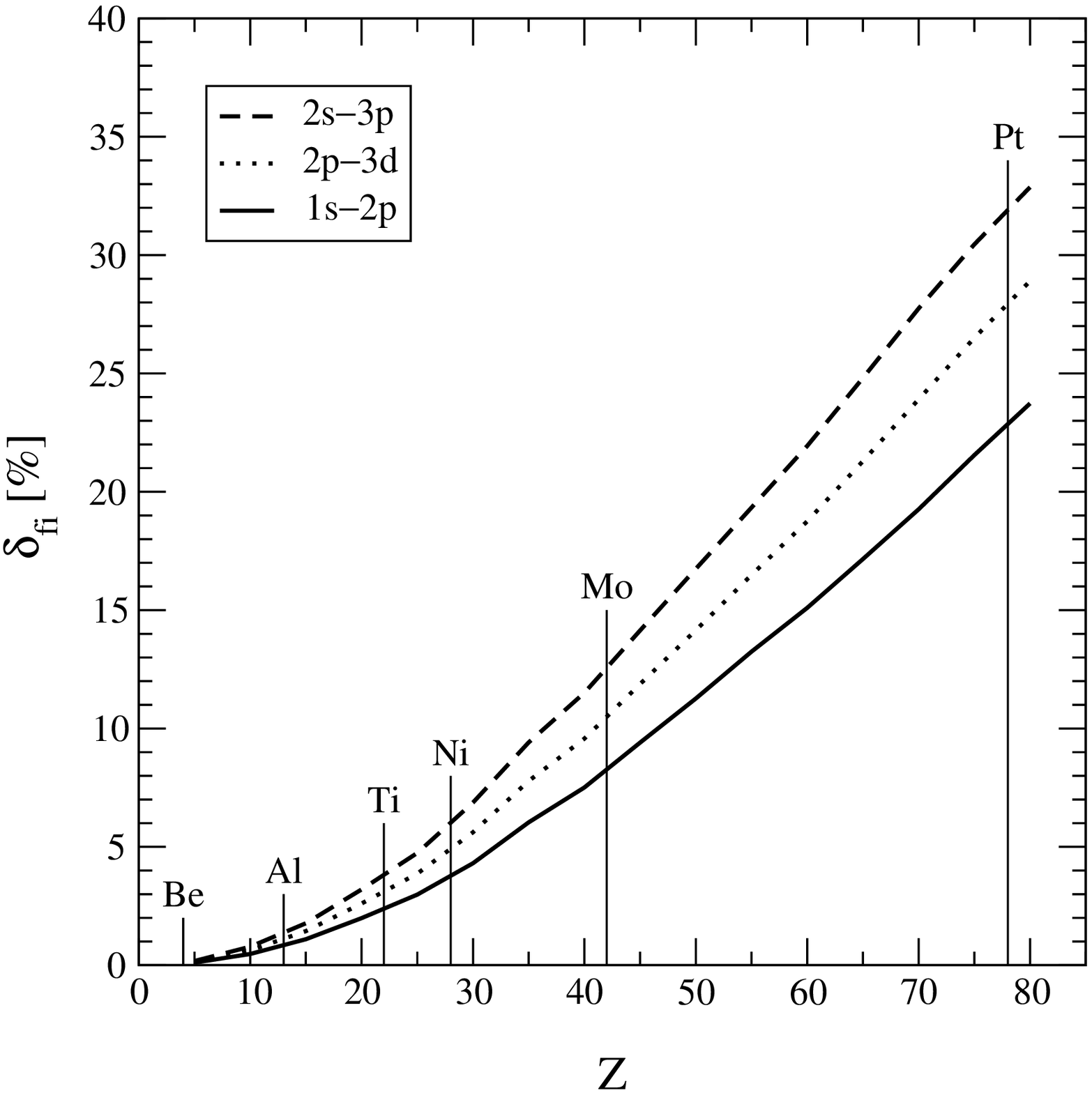}}
\caption{The relative difference between the lowest order and
  the Glauber results 
$\delta_{fi}=(\sigma_{LO}-\sigma_{GL})/\sigma_{LO}$
is shown for different transitions as a function of the charge number of the 
target.}
\label{figglauber}
\end{minipage}
\hfill
\begin{minipage}[b]{0.48\hsize}
\leavevmode
\centerline{\includegraphics[width=4.0cm]{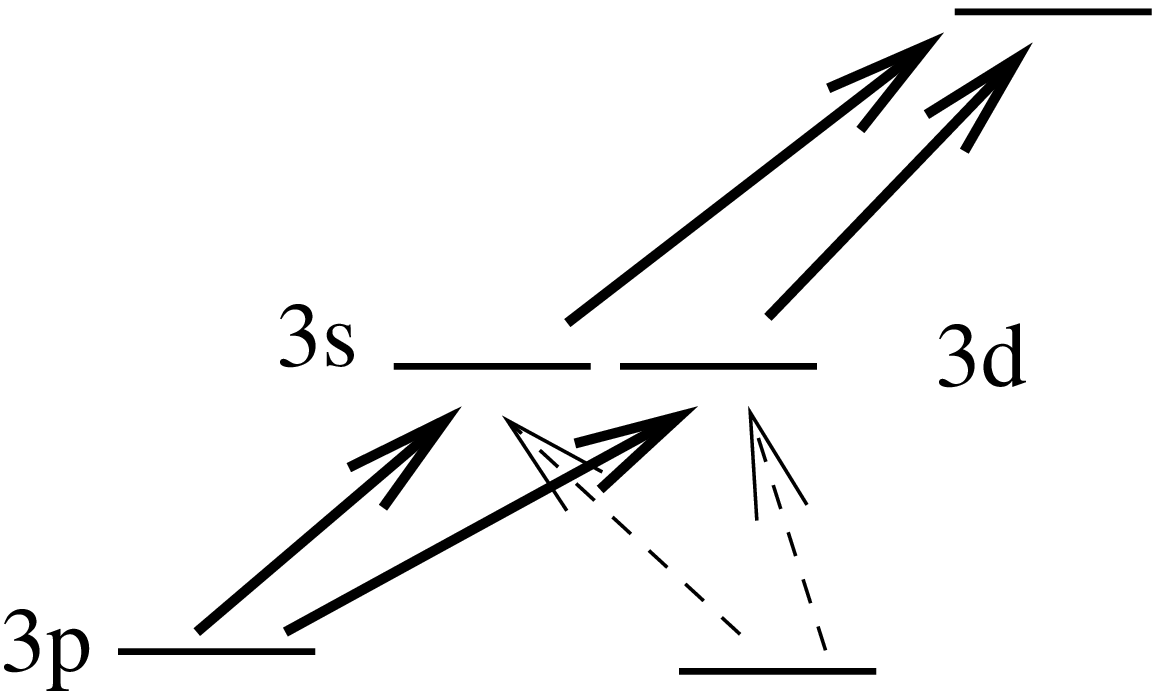}
\includegraphics[width=4.0cm]{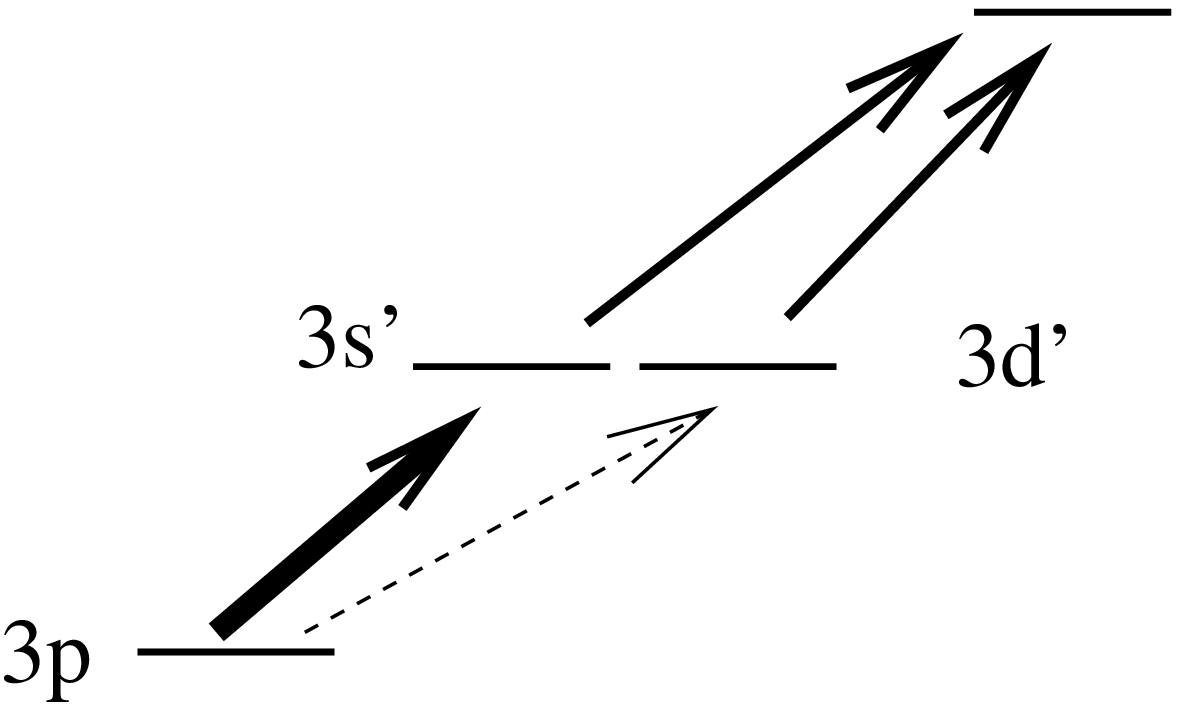}}
\caption{In the optimal mixture approach a linear combination (of the
$3s$ and $3d$ ($m=0$) in the case depicted here) is chosen, which maximizes
the transition to one of the new set of states (``$3s'$'') from the dominant
feeding state, which in this case is $3p$, $m=\pm 1$. The propagation of
pionium through matter is then done employing this new set of states.}
\label{figoptimal}
\end{minipage}
\end{figure}

\begin{figure}[ht]
\centerline{\includegraphics[width=5.0cm]{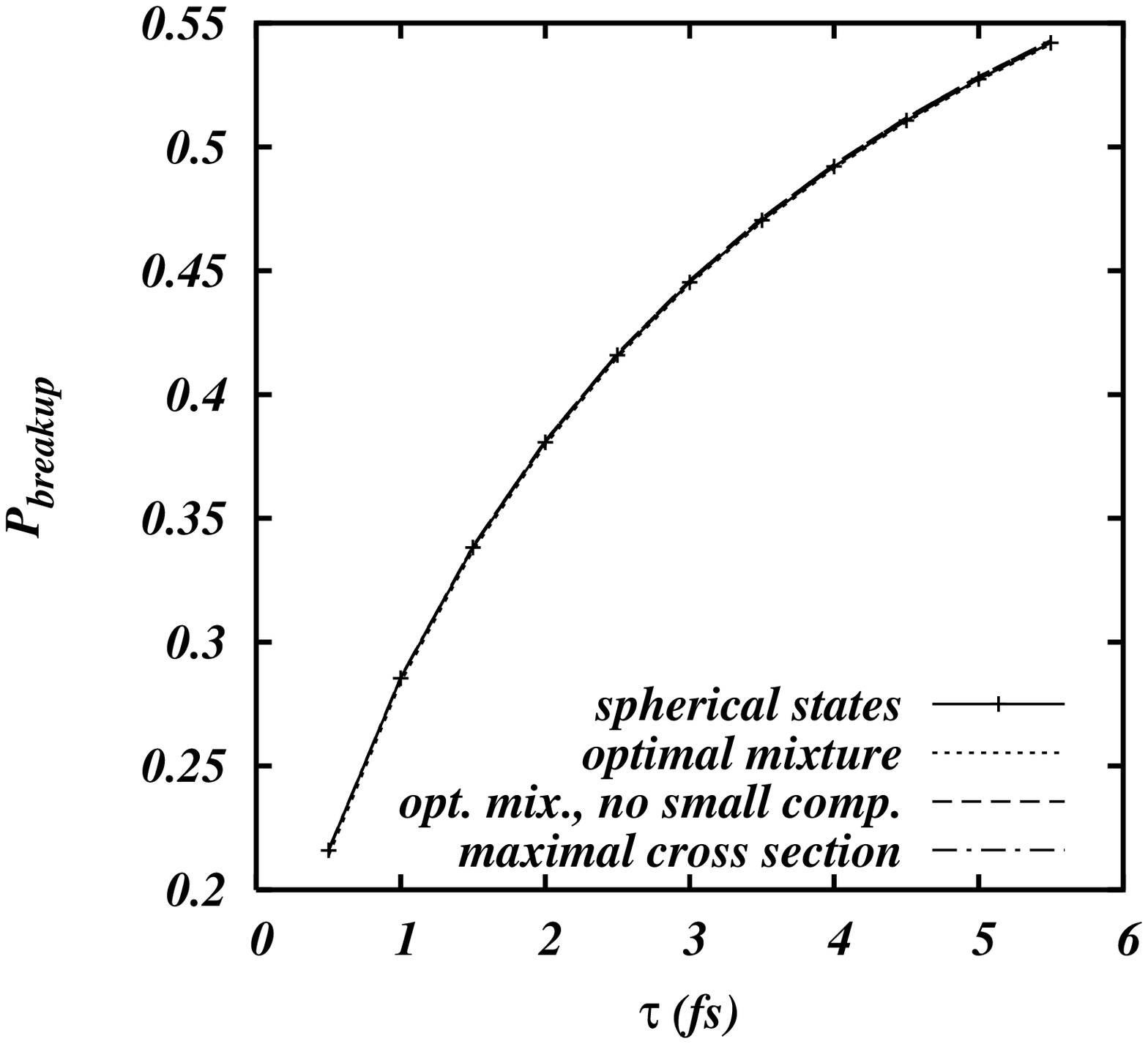}
(A)\hspace{2.8cm}
\includegraphics[width=5.0cm]{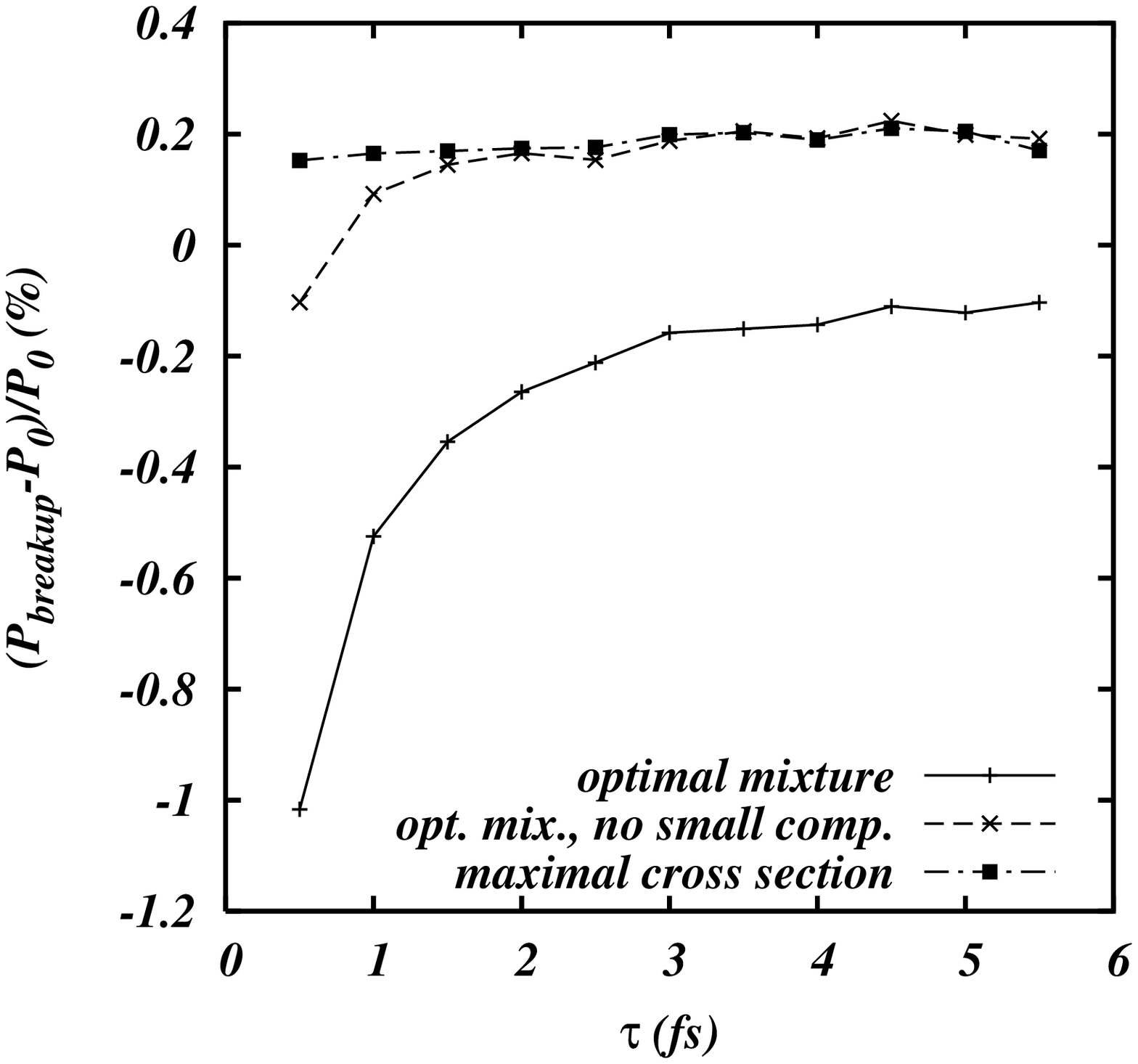}
(B)}
\caption{In (A) the change of the breakup ratio $P_{\mathrm{br}}$ as a
function of the pionium lifetime $\tau$ is shown for the four different 
models, see the text for detail. (B) shows the relative difference of the
three new models compared to the one using spherical states.
In the calculation, a 95 $\mu$m Ni target 
and a pionium momentum of $p=4.7\ \mbox{GeV}/c$ were used.}
\label{figmixed}
\end{figure}

In order to achieve the required accuracy of 1\% for the cross section needed
for the analysis, a number of effects need to be incorporated in the 
calculations. Starting from the Born cross section within the semiclassical
approach, target inelastic effects, magnetic terms and relativistic 
corrections have been calculated and included in the tabulation of the cross
section, where needed, see \citetwo{Schumann:2002xx}{Santamarina:2003ns} 
and references therein.
Within a Glauber calculation it was found that the higher order photon exchange
is important, as the excitation cross sections change relative to the lowest 
order results by up to 20\%, see Fig.~\ref{figglauber}. An independent test of 
the Glauber results by using a coupled channel approach for some important 
transitions is currently under way. We have found already that transitions to 
the continuum are important and cannot be neglected. We are including them 
either within a discretization of the continuum (``Weyl states'') or
by a perturbative coupling to the continuum only.

It has become clear in the meantime, that the density formalism used to 
describe the propagation of pionium through the target needs to be tested 
more thoroughly. By choosing two different decompositions 
(spherical and parabolic coordinates) for degenerate states, it was pointed 
out in \cite{Afanasyev:1999xx} that changes of the predicted life time
of the order of a few percent can be found. More recently a full quantum
mechanical approach to the propagation within a density matrix approach has 
been proposed \cite{Voskresenskaya:2002xx}. 
We have employed an ``optimal mixture'' approach. The idea is to stay
within the density formalism, but to use optimal linear combinations
of degenerate states that
maximize the transition from the dominant feeding states into them, see 
Fig.~\ref{figoptimal}. Due to $m$-number conservation and $z$-parity, the 
important cases are the $3s/3d$, $4s/4d$ ($m=0$) and the $4p/4f$ ($m=\pm 1$) 
states. The optimal linear combination was found for a mixing angle 
of about 161$^o$ ($3s/3d$), 155$^o$ ($4s/4d$) and 170$^o$ ($4p/4f$). The
resulting new basis states are therefore close to the spherical
one. Using these new states within 
the density formalism, and adapting production and annihilation cross
sections to 
incorporate their mixed nature, we find a deviation from the pure spherical
basis of about 0.2\%, see Fig.~\ref{figmixed}. A test of the possible error
within this approach was made by dropping the transitions to the 
``small component'' altogether, which gives a change of about -0.5\%.
As a different approach we assume that the whole cross section to the 
degenerate states goes to $s$ or $p$ states only. With this we find
good agreement with the ``optimal mixture'' model. From this we conclude,
that the accuracy of our approach is of the order of 0.5\% with a deviation
from the density formalism up to now of 0.2\%.

\newabstract 
\label{abs:Afanasyev}

\begin{center}
  {\large\bf Production of the long-lived excited states of $\pi^+\pi^-$
    atoms in view to perform an experiment on their observation and
    measurement the Lamb shift}\\[0.5cm]
{\bf L. Afanasyev} and L. Nemenov\\[0.3cm]
Joint Institute for Nuclear Research, 141980 Dubna,Moscow Region, Russia
\end{center}

Measurement of the $\pi^+\pi^-$ atom lifetime in the DIRAC experiment
will allows one to obtain in the model independent way the value
$|a_0-a_2|$, the difference of the $s$-wave $\pi\pi$ scattering
lengths with the isotop spin 0 and 2 correspondently \cite{dirac}. To
get the values of $a_0$ and $a_2$ separately basing on the
$\pi^+\pi^-$ atom data, one may use the fact that the energy splitting
between levels $ns$ and $np$, $\Delta E_n=E_{ns}-E_{np}$, depends on the
the another combination of scattering lengths: $2a_0+a_2$. Thus the
measurement the energy splitting coupled with the lifetime measurement
would provide a determination of $a_0$ and $a_2$ separately
\citetwo{dirac}{nem85}.

The lifetimes of the $np$ states are significantly, 3--5 order, higher
in compare with the ground state \cite{nem85}. For that reason atoms
in $np$ states have the mean paths of teens centimeters. Methods of
$\Delta E_n$ measurement proposed in papers \cite{nem01} and
\cite{nem02} based on observation of the interference between $ns$ and
$np(m=0)$ states in the external electro-magnetic fields.

Production of $\pi^+\pi^-$ atom in the $np$ states have been
considered for different target materials and thicknesses with the
intent to optimize the experimental conditions for their
observation. It have been shown that for the DIRAC experiment usage of
thiner targets with smaller $Z$ provides increase of yield of $np$
states and a better ratio to the atom break-up. In the following table
a set of targets providing the highest yield of  $\pi^+\pi^-$ atom
states with the magnetic quantum number $l\ge1$ are shown.

\vspace{1mm}
\noindent \begin{tabular}{|c|c|c|c|c|c|c|c|}
\hline
Target & Thickness & Break-up & $\sum (l\ge1)$ & $2p_0$ & $3p_0$ & $4p_0$ &  $\sum(l=1,m=0)$\\
Z & $\mu m$ &&&&&& \\
\hline
04 & 50 & 2.63\% & 5.86\% & 1.05\% & 0.54\% & 0.20\% & 1.93\% \\
06 & 50 & 5.00\% & 6.92\% & 1.46\% & 0.51\% & 0.16\% & 2.52\% \\
13 & 20 & 5.28\% & 7.84\% & 1.75\% & 0.57\% & 0.18\% & 2.63\% \\
28 & 5  & 9.42\% & 9.69\% & 2.40\% & 0.58\% & 0.18\% & 3.29\% \\
78 & 2  & 18.8\% & 10.5\% & 2.70\% & 0.55\% & 0.16\% & 3.53\% \\
\hline
\end{tabular}
\vspace{1mm}
 
The experimental data already collected in the DIRAC experiment with
single and multi-layer nickel targets confirm existence of the excited
states of $\pi^+\pi^-$ atom with the mean paths much longer than 1~mm.

\newabstract 
\label{abs:Tarasov2}

\begin{center}
  {\large\bf The quantum-mechanical treatment of pionium
    internal dynamics in matter}\\[0.5cm]
  {\bf A. Tarasov} and O. Voskresenskaya\\[0.3cm]
  Joint Institute for Nuclear Research, 141980 Dubna,Moscow Region,
  Russia
\end{center}

The quantum-mechanical consideration of a passage of fast dimesoatoms
through matter is given. A set of quantum-kinetic equations for the
density matrix elements describing their internal state evolution is
derived. It is shown that probabilistic description of internal
dynamics of hydrogen-like atoms is impossible even at sufficiently low
energies because of the ``accidental'' degeneracy of their energy
levels.

\newabstract 
\label{abs:Ivanov}

\begin{center}
{\large\bf On radiative decay channels of pionic and kaonic
hydrogen}\\[0.5cm] {\bf A. N. Ivanov}, M. Cargnelli, M. Faber,
A. Hirtl, J. Marton, N. I. Troitskaya, and J. Zmeskal\\[0.3cm]
Atominstitute, Vienna University of Technology and Institute of Medium
Energy Physics, Austrian Academy of Sciences, Vienna, Austria
\\[0.3cm]
\end{center}

In \cite{IV1} we have developed a quantum field theoretic,
relativistic covariant and model--independent approach to the
description of the energy level displacements of the ground and
excited $n\ell$ states of pionic and kaonic hydrogen $A_{h^-p}$, where
$h^- = \pi^-$ or $K^-$. In order to estimate of the contribution of
the radiative decay channels of pionic and kaonic hydrogen we apply a
soft--pion and soft--kaon technique \cite{IV1} (nucl--th/0310081). For
the Panofsky ratio we get the result \cite{IV1} (nucl--th/0310081)
\begin{eqnarray}\label{label1}
\hspace{-0.3in}\frac{1}{P} = \frac{\sigma(\pi^-p \to
n\gamma)}{\sigma(\pi^-p \to n\pi^0)} = \frac{\Gamma_{1s}(A_{\pi^-p}
\to n\gamma)}{\Gamma_{1s}(A_{\pi^-p} \to n\pi^0)} = 0.681\pm 0.048
\end{eqnarray}
agreeing well with the experimental value $1/P_{\exp} = 0.647 \pm
0.004$ \cite{JS77}. As it is shown in \cite{TE03} the agreement with
the experimental data can be improved accounting for the form factor
of the axial--vector current, the matrix element of which defines the
amplitude of the reaction $\pi^-p \to n\gamma$ in the soft--pion
limit, i.e. at leading order of ChPT (Gasser $\&$ Leutwyler). For the
Panofsky ratio of kaonic hydrogen we get \cite{IV1} (nucl--th/0310081)
\begin{eqnarray}\label{label2}
\frac{1}{P} = \frac{\Gamma_{1s}(A_{K^-p} \to \Lambda^0\gamma) +
\Gamma_{1s}(A_{K^-p} \to \Sigma^0\gamma)}{\Gamma_{1s}} = (3.97 \pm
0.47)\times 10^{-3},
\end{eqnarray}
where $\Gamma(A_{K^-p} \to \Lambda^0\gamma) = (0.82\pm 0.04)\,{\rm
eV}$, $\Gamma_{1s}(A_{K^-p} \to \Sigma^0\gamma) = (0.08\pm 0.01)\,{\rm
eV}$ and $\Gamma_{1s} = (227 \pm 27)\,{\rm eV}$. Thus, the
contributions of the radiative decay channels of kaonic hydrogen can
be neglected for the theoretical analysis of experimental data on the
energy level displacement of the ground state of kaonic hydrogen by
the DEAR Collaboration: $\epsilon^{\exp}_{1s} = (183\pm 62)\,{\rm eV}$
and $\Gamma^{\exp}_{1s} = (213 \pm 138)\,{\rm eV}$ (see M. Cargnelli,
{\it DEAR--Kaonic Hydrogen: First Results}, hep--ph/0311212, p.125).

\newabstract 
\label{abs:Zemp}

\begin{center}
{\large\bf Deser-type formula for pionic hydrogen}\\[0.5cm]
Peter Zemp\\[0.3cm]
ITP, University of Bern, Sidlerstrasse 5, 3012 Bern, Switzerland\\[0.1cm]
\end{center}

The decay width of the ground state of pionic hydrogen is dominated by
the two decay channels $(\pi^-p)_{1s}\to \pi^0n$ and $(\pi^-p)_{1s}\to
n\gamma$.  Using $\delta$ as a common counting for the light quark mass
difference $m_u-m_d$ and for the fine-structure constant $\alpha$, the leading
term of the former (latter) decay channel is of order $\delta^{7/2}$
$(\delta^4)$. The decay width through these two channels can be expressed by a
Deser-type formula
\begin{equation}
  \label{eq:deser-type}
  \Gamma_{1s}= 4\, \alpha^3 M_r^2  q_{\scriptscriptstyle 0} \left ( 1 +
    \frac{1}{P} \right ) [a_{\pi^-p \to \pi^0
    n}(1+\delta_{\scriptscriptstyle\Gamma})]^2 \;.
\end{equation}
Since the Panofsky ratio $P = \sigma(\pi^-p \to \pi^0 n) / \sigma(\pi^- p \to
n \gamma)$ is about $1.5$, the $n \gamma$ channel amounts to a 60\%
correction. To the best of my knowledge, this formula was first established in
the framework of potential models \citetwo{Rasche:mp}{Sigg:qd}. I want to verify
its validity in QCD~+~QED, using the technique of non-relativistic effective
field theories \citethree{Caswell:1985ui}{Antonelli:2000mc}{Lyubovitskij:2000kk}.  Constructing the effective Lagrangian and
solving the master equation, one encounters the obstacle that the photon in
the intermediate state carries a hard momentum of order $M_{\pi}$.  To
circumvent this problem, I do not incorporate $n\gamma$ intermediate states
with effective fields. Instead I introduce a non-hermitian contact term
\cite{Caswell:1985ui}
(cf.~also~\cite{Gasser:2001un}) in the Lagrangian with the coupling
$d_1^R + i d_1^I$,
\begin{equation}
  \label{eq:intLagrangian}
  \mathcal{L}^{n \gamma}_{\rm I} = (d_1^R + i d_1^I) \; \psi^\dagger(x)
  \psi(x) \; \pi^\dagger_-(x) \pi_-(x)\;.
\end{equation}
The imaginary part $d_1^I$ replaces the imaginary part generated by the
intermediate $n\gamma$ state.  With this contact term, I can prove that
formula (\ref{eq:deser-type}) is valid at order $\delta^4$. It
remains\footnote{P.~Zemp, work in progress.} to evaluate $\Gamma_{1s}$ at
order $\delta^{9/2}$ and to determine $\delta_{\scriptscriptstyle \Gamma}$ in
equation (\ref{eq:deser-type}).

\newabstract 
\label{abs:Ericson}

\begin{center}
{\large\bf Electromagnetic corrections to  scattering lengths from hydrogenic atoms applied to the $\pi ^-$p system}\\[0.5cm]
{\bf T.E.O.~Ericson}$^1$,  B.~Loiseau$^2$, and S.~Wycech $^3$\\[0.3cm]
$^1$CERN, CH-1211 Geneva 23, Switzerland, and 
 TSL, Box 533, S-75121 Uppsala, Sweden\\[0.1cm]
$^2$LPNHE, Universit\'{e} P. \& M. Curie,
 4 Place Jussieu, F-75252 Paris, France\\[0.1cm]
$^3$Soltan Institute for Nuclear Studies, PL-00681 Warszawa,
 Poland
\end{center}

 Motivated by the accurate measurements of
 strong interaction 1s shifts and width in pionic hydrogen, we explore the 
atomic corrections  to the leading order Born relation of the (complex) scattering length $a^h$ 
 to the 
(complex) strong interaction level shift $\epsilon _{1s}-i\Gamma _{1s}/2$.
In terms of the    reduced mass  $\mu $ and the deviation $\delta _{1s}$ from the Born relation:
\begin{equation}
\epsilon _{1s}=-\frac {4\pi }{2\mu }~\Phi _B(0)^2~{\rm Re} \ a^h(1+\delta _{1s})=
4\mu \alpha ~{\rm Re } \ a^h(1+\delta _{1s})~E_B~,
\end{equation}
where $\Phi _B(r)$ and $E_B=-\mu \alpha ^2/2$ are the Bohr wave function  and 
 Bohr energy, respectively. 
\newline
 To display the
 physical mechanisms at work, we 
first consider an  oversimplified model with the merit of an explicit solution.
 It  is later realistically generalized~\cite{Ericson:2003ju}.
 The 
hadronic scattering length 
 $a^h$ is taken to result from a zero range interaction. The charge distribution is approximated 
by concentrating the charge to a 
spherical shell of radius $R$ with the scattering interaction at its center. 
The solution to order $\alpha ^2\ln \alpha $ for a single channel
is:
\begin{equation}
\delta _{1s}  =   
 -2\frac {R}{r_B} 
  +2\frac {a^h}{r_B}\left[2-\gamma -\log (2\alpha mR)\right]
  \label{deltapoint2}=\delta _{1s,1}+\delta _{1s,2}.
\end{equation}
\newline
An energy dependent amplitude  with a threshold  expansion 
$a^h(q^2)=a^h+b^hq^2+..$ gives the additional term $\delta _{1s,3}=(2\mu \alpha /R){\rm Re} b^h/{\rm Re}a^h$.
These three generic contributions  have simple physical interpretations.
The  linear term in R represents the change of the wave function at the origin of the extended charge with
$\delta \Phi (0)/\Phi _{B}(0)=-\alpha \mu R$; the  term proportional to $1/R$ expresses that
the relevant kinetic energy is the depth of the extended charge Coulomb potential and  not the small binding energy.
The  term proportional to the scattering length $\delta _{1s,2}$ is a renormalization induced by consistency of the 
strong energy shift with the scattering length producing it. 
 A realistic generalization   to  the
$\pi ^-p \rightarrow \pi ^-p$ channel in the  coupled  channel $\pi $N  system
with the empirical charge distribution and mass difference     gives in percent:
$
\delta _{1s,1}=-0.853(8)  ;\delta _{1s,2}=0.701(4)   ; \delta _{1s,3}= -0.95(29) ;\delta _{1s}=-0.62(29).$
These results correspond to an interpretation of the energy shift in terms of a scattering length
in the absence of an external, removable Coulomb field.
We have made no corrections for the 
internal e.m. contributions to the masses.
 Our relation cannot therefore  be readily compared to  the
EFT expansion based on the QCD Lagrangian \cite {Gasser02}.

\newabstract 
\label{abs:Gotta}

\begin{center}
{\large\bf Ground-State Shift in Pionic hydrogen}\\[0.5cm]
Detlev Gotta {\it for the PIONIC HYDROGEN collaboration}\\[0.3cm]
Institut f\"ur Kernphysik, Forschungszentrum J\"ulich, D-52425 J\"ulich\\[0.3cm]

\end{center}

The measurement of ground--state transitions in pionic hydrogen allows to  
determine the isoscalar and isovector scattering lengths $a^{+}$ and $a^{-}$, which 
describe the $\pi$N s--wave interaction\,\citethree{Lyu00}{Gas03}{Zem03}. In addition, 
the $\pi$N coupling constant can be extracted by applying the Goldhaber--Miyazawa--Oehme 
sum rule\,\citetwo{Gol55}{Eri02}. To improve on the accuracy achieved by previous 
measurements\,\cite{Sch01}, a thorough study of the atomic cascade has been started 
at the Paul--Scherrer--Institut using the new cyclotron trap, a cryogenic target, and a 
Bragg spectrometer equipped with spherically bent silicon and quartz crystals and a 
large--area CCD array\,\cite{Nel02} (PSI experiment R--98.01\,\citetwo{R98.01}{Ana03}) . 

At first, the possibility of radiative de--excitation of the $\pi$H atom -- when 
bound into complex molecules formed during collisions 
$\pi^{-} p+H_{2}\rightarrow [(pp\pi^{-})p]ee$\,\cite{Jon99} -- was studied by searching for a
density dependence of the $\pi$H(3p-1s) transition energy. No density effect was found in 
a pressure range from 3.5\,bar to liquid and, consequently, the measured line shift $\epsilon_{1s}$ 
can be attributed exclusively to the strong interaction. The value of $\epsilon_{1s} =7.120\pm 0.013$\,eV 
(preliminary) measured in this experiment\,\citetwo{Ana03}{Hen03} was found to be in good agreement with 
the result of the previous experiment\,\cite{Sch01}. The experimental error was improved by more than a 
factor of two. The precise value for $\epsilon_{1s}$ from $\pi$D, which has been used for the 
determination of $a^{+}$ and $a^{-}$ because of the poor knowledge of $\Gamma_{1s}$\,\cite{Eri02}, 
may be questioned because radiative de--excitation should be enhanced considerably.

At present, the accuracy for $\Gamma_{1s}$ (7\%) is limited by the not sufficiently well 
known correction for the Doppler broadening of the X--ray lines\,\cite{Sch01}, which is 
is caused by conversion of de--excitation energy into kinetic energy 
during collisions (Coulomb de--excitation)\,\cite{Bad01}. The efforts in the extraction 
of the strong--interaction broadening are dicussed by L.~Simons\,\cite{SimHad03}.

\newabstract 
\label{abs:Jensen}

\begin{center}
{\large\bf Cascade model predictions: $K^-p$, $K^-d$, and $\pi^-p$}\\[0.5cm]
T.S. Jensen\\[0.3cm]
LKB, 
Ecole Normale Sup\'erieure et Universit\'e Pierre et Marie Curie,
Case 74,\\
 4 Place Jussieu,
F-75252 Paris Cedex 05,
France

\end{center}

The extended standard cascade model \cite{jensen02} has been used to study the
atomic cascade in exotic hydrogen atoms.
The model is based on cross sections for the collisional
processes: Stark transitions, elastic scattering, nuclear absorption during
collisions, Coulomb deexcitation, and external Auger effect. 
The X-ray yields, cascade times, and kinetic energy distributions in exotic hydrogen
atoms have been calculated and compared with experimental data. 
Cascade model predictions are important for the analysis of experimental data and for
planning future experiments. The following three experiments are of particular interest:

\begin{itemize}

\item{
The DEAR Collaboration has measured the $K$ X-ray spectrum in 
kaonic hydrogen and results for the $1s$ shift/width and the X-ray yields are in 
progress \cite{cargnelli03}. The $K$ yields depend strongly on the $2p$ strong interaction 
width so cascade model predictions can be used to determine it.  
}

\item{
The SIDDHARTA Collaboration plans to measure the $K$ X-ray spectrum
in kaonic deuterium
from which the $1s$ shift/width can be extracted \cite{petrascu03}. The experiment
depends crucially on the $K$ X-ray yield which is reduced by absorption from the 
$ns$ and $np$ states. Our cascade calculations show that absorption from $ns$ states 
does not make X-ray measurements infeasible. 
The poorly known $np$ state absorption could, however, reduce the yields significantly. 
}

\item{
The pionic hydrogen $1s$ shift/width experiment at PSI is in 
progress \cite{psiexp}. In order to extract the strong interaction width from the measured
line profiles corrections due the  Doppler broadening must be taken into account--{\it i.e.}
the kinetic energy distributions at the instant of the radiative transitions are needed. 
At present, the calculated cross sections which are used as input by the cascade program 
are not accurate enough for an {\it ab initio} calculation of the Doppler correction to
the required precision. Until improved results for the collisional processes become
available a combination of cascade model constraints and
a fitting procedure is the best option.
}

\end{itemize}

\noindent  Acknowledgment: This work was supported by the Swiss National Science Foundation.

\newabstract 
\label{abs:Simons}

\begin{center}
{\large\bf Experimental determination of the strong interaction ground state width in pionic hydrogen}\\[0.5cm]
L. M. Simons, for the PIONIC HYDROGEN collaboration. \\[0.3cm]
 Paul Scherrer Institute, CH 5232 Villigen PSI, Switzerland\\[0.1cm]

\end{center}

The importance of the measurement of the ground state shift and width 
from a measurement of X-rays from pionic hydrogen atoms has been addressed 
at this workshop by D. Gotta \cite{gotta}. Whereas first results could be obtained 
for the shift from recent measurements the extraction of the width is still 
subject of investigations. Two main difficulties are withstanding 
a direct extraction of the width. The first difficulty is inherent to the 
method of the measurement as a crystal spectrometer is used in
 order to achieve 
high resolution. The response function of this device is a complicated 
 function of crystal and imaging properties  and has to be 
determined in an independent measurement.
Such a measurement has been performed in the meanwhile by using an 
ECR (electron cyclotron resonance) source \citetwo{biri}{dimis}. This source 
 produces an extended 
source of quasi-monoenergetic X-rays in the energy range of the 
pionic hydrogen X-rays. Astonishingly the measurements showed 
no sizeable deviation of the resolution function from 
the theoretical values \cite{XOP}. This fact allows to produce 
response functions for each of the transition energies and crystal
 types ( $\alpha$ quartz(10-1) and Si(111)) as well and served as 
a basis for the analysis of the measured spectra.

The second not yet solved difficulty is caused by the fact that 
 the pionic hydrogen atoms 
are not at rest at the instant of X-ray emission. This  leads to a Doppler 
broadening which renders the extraction of the strong 
interaction width to be very difficult. The theory of the different 
processes changing the state of the velocity of the pionic hydrogen atoms 
 is discussed by T. Jensen in a contribution to this 
workshop \cite{jensen}.   Especially kinetic energies of about some 10 eV
will result in a Doppler broadening faking a strong interaction 
broadening of about 900 meV. A thorough understanding of the 
acceleration processes is therefore indispensable to reach the goal 
of an accuracy of about 1$\%$ required in the proposal \cite{proposal}.
From a preliminary evaluation of the 3-1 and the 4-1 transition, however,
 even now 
a safe upper limit for the strong interaction broadening of 850 meV can 
be extracted.

\newabstract 
\label{abs:Sazdjian}

\begin{center}
{\large\bf Calculating $\mbox{\boldmath$\pi$}\mathbf{K}$ atom properties\\
in the constraint theory approach}\\[0.5cm]
H. Sazdjian\\[0.3cm]
Groupe de Physique Th\'eorique, Institut de Physique Nucl\'eaire,\\
Universit\'e Paris XI, 91406 Orsay Cedex, France\\
\small{E-mail: sazdjian@ipno.in2p3.fr}
\end{center}

The constraint theory approach reduces the Bethe--Salpeter equation to a
covariant three-dimensional equation. It has been applied recently to the
calculation of the pionium lifetime with its relativistic corrections. The
method can also be applied to the evaluation of the lifetime and energy
splittings of excited states of the $\pi K$ hadronic atoms, for the
observation of which experimental projects are being prepared.
The calculations can be done in covariant arbitrary gauges for the photon
propagator present in the QED type diagrams that contribute to the
evaluation of part of the electromagnetic corrections. Observable
quantities should be independent of the gauge parameter. Cancellation of
infra-red singularities between various off-mass shell scattering diagrams
and effective three-dimensional diagrams, resulting from the
three-dimensional reduction procedure, should occur in order to render the
interaction potentials free of such singularities. In that case, the 
potentials that are present in the bound state equation can be analytically
continued to the $\pi K$ threshold and identified with the real parts of 
the infra-red regularized on-mass shell $\pi K$ scattering amplitudes at 
threshold, including isospin breaking and electromagnetism at leading order.
The expressions of those amplitudes can then be used for the evaluation of
the various physical quantities.

\newabstract 
\label{abs:Gasser}

\begin{center}
{\large\bf On the precision of the theoretical predictions for
  $\pi\pi$ scattering}\\[0.5cm]
J. Gasser\\[.5cm]
Institute for Theoretical Physics, University of Bern, 
Sidlerstrasse 5, 3012 Bern, Switzerland\\[0.1cm]
\end{center}

In two recent papers \cite{py}, Pel\'aez and 
Yndur\'ain evaluate 
some of the low energy
observables of $\pi\pi$ scattering and obtain flat 
disagreement with our earlier
results \cite{pipicgl}. The authors work with unsubtracted 
dispersion relations,
so that the outcome of their calculation is very 
sensitive to the poorly 
known high energy
behaviour of the scattering amplitude. They claim that 
the asymptotic
representation we used in \citetwo{pipicgl}{physrep} is incorrect 
and propose an alternative one. 
We have repeated \cite{pipica} their calculations on the 
basis of the standard, 
subtracted
fixed-$t$ dispersion relations, using their asymptotics. 
The outcome fully
confirms our earlier findings. Moreover, we show that the
Regge parametrization proposed by these authors for the 
region above 1.4 GeV
violates crossing symmetry: Their ansatz is not consistent 
with the
behaviour observed at low energies.

\newabstract 
\label{abs:Schweizer}

\begin{center}
{\large\bf  Energy shift and decay width of the $\pi K$ atom}\\[0.5cm]
J. Schweizer\\[0.3cm]
ITP, University of Bern, Sidlerstrasse 5, 3012 Bern, Switzerland
\end{center}
\small{Nearly fifty years ago, Deser et al. \cite{Deser:1954vq} derived the
formula for the width of pionic hydrogen at leading order in isospin symmetry
breaking. Similar formulas also hold for pionium and the $\pi^- K^+$ atom. These
Deser-type relations allow to extract the scattering lengths from 
measurements of the decay width and the strong energy shift. The DIRAC collaboration
at CERN \cite{Adeva:1994xz} aims to measure the pionium lifetime to
$10\%$ accuracy which allows to determine the S-wave $\pi\pi$
scattering lengths difference $|a_0^0-a_0^2|$ at $5\%$ precision. New
experiments are proposed at CERN PS and J-PARC in Japan \cite{Nemenov}. 

To determine the scattering lengths from such precision measurements, the theoretical expressions must be known to an accuracy that matches the experimental
precision. We evaluated the width and the energy shift for the ground state of
the $\pi^- K^+$ atom within the non-relativistic effective Lagrangian
framework \cite{Caswell:1985ui}. The result reads at next-to-leading order in isospin symmetry
breaking, where both
$\alpha\simeq1/137$ and $m_u-m_d$ count as small quantities of order $\delta$,
\begin{equation}
  \Gamma_{\pi^0 K^0} =
  8\alpha^3\mu_+^2p^*\mathcal{A}^2\left(1+K\right), \quad\mathcal{A} = -\frac{1}{8\sqrt{2}\pi}\frac{1}{M_{\pi^+}+M_{K^+}}{\rm
  Re}\,A^{00;\pm}_{\rm thr}+{\it o}(\delta),
\label{Gammapi0K0}
\end{equation}
with 
\begin{eqnarray}
p^* &=& \left[\frac{M_{\pi^+}
  \Delta_K+M_{K^+}\Delta_\pi}{M_{\pi^+}+M_{K^+}}-\alpha^2\mu_+^2+\frac{(\Delta_K-\Delta_\pi)^2}{4(M_{\pi^+}+M_{K^+})^2}\right]^{\frac{1}{2}},\nonumber\\
K &=& \frac{M_{\pi^+}
  \Delta_K+M_{K^+}\Delta_\pi}{M_{\pi^+}+M_{K^+}}{a^+_0}^2-4\alpha\mu_+\left[{\rm
  ln}\alpha-1\right](a^+_0+a^-_0)+{\it o}(\delta).
\end{eqnarray}
Here $\Delta_\pi =
M_{\pi^+}^2-M_{\pi^0}^2$, $\Delta_K =
M_{K^+}^2-M_{K^0}^2$ and $\mu_+$ denotes the charged reduced mass. The quantity ${\rm Re}\,A^{00;\pm}_{\rm thr}$ is
determined as follows. One evaluates the relativistic $\pi^- K^+ \rightarrow
\pi^0 K^0$ amplitude at order $\delta$ near threshold, see
Refs.~\citetwo{Kubis:2001ij}{Nehme:2001wf}. The real part of this
matrix element contains a
singularity $\sim 1/|\bf{p}|$ at threshold ($\bf{p}$ denotes the c.m. momentum of the charged pion and kaon). The constant term in the
threshold expansion corresponds to ${\rm Re}\,A^{00;\pm}_{\rm thr}$. Further,
$\mathcal{A}$ is normalized such that in
the isospin symmetry limit it coincides with the isospin odd scattering length
$a_0^-$. The isospin even and odd $\pi K$ scattering lengths\footnote{We use
  the same notation as in Ref. \cite{Kubis:2001ij}.} $a_0^+$ and
$a_0^-$ are defined in
QCD at $m_u = m_d$ and $M_\pi \doteq M_{\pi^+}$, $M_K \doteq
M_{K^+}$.
}

\newabstract 
\label{abs:Buettiker}

{\setlength{\parindent}{0mm}
\begin{center}
{\large\bf Roy--Steiner equations for \boldmath{$\pi K$} scattering}\\[5mm]
{\bf Paul B\"uttiker}$^1$, Sebastien Descotes--Genon$^2$, Bachir
Moussallam$^3$\\[3mm]
$^1$ HISKP, Universit\"at Bonn, D--53115 Bonn,\\[0.1cm]
$^2$ LPT, Universit{\'e} Paris--Sud, F-91406 Orsay,\\[0.1cm]
$^3$ IPN, Universit{\'e} Paris--Sud, F-91406 Orsay
\end{center}
$\pi K$ scattering is the most simple $SU(3)$--process involving
strange quarks. It is therefore an ideal place to test chiral
predictions with non--vanishing strangeness. While this process is
interesting by itself, a detailed knowledge of $\pi K$ 
scattering also contributes to the understanding of the flavour
dependence of the order parameters of Chiral Perturbation Theory
(ChPT) \cite{stern}.
%
In our work we use Roy--Steiner equations  to analyze the available
experimental $\pi K$ and $\pi\pi\to K\bar{K}$ data and to explore the
low energy region where no such data are available \cite{ABM}.
\begin{floatingfigure}{7.0cm}
\includegraphics[scale=0.55]{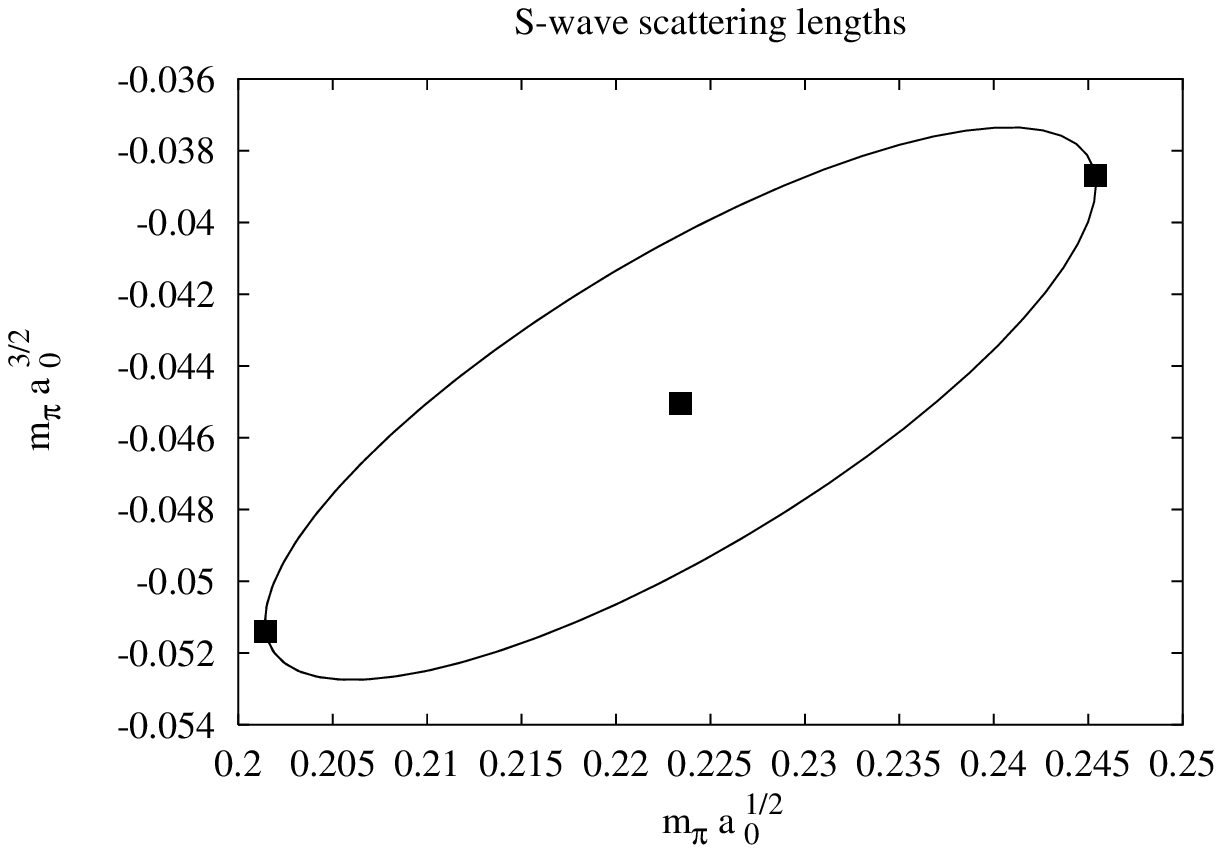}\\
Fig:~$1$--$\sigma$ error ellipse for the scattering\\\phantom{Fig:~}lengths
  $a^{1/2}_0$ and $a^{3/2}_0$.
\end{floatingfigure}
Taking the higher partial waves and all $S$-- and $P$--waves above
the energy region from threshold to $\approx$ 1 GeV. Our solutions for
the low--energy region turn out to be in general poor
agreement with the available experimental data.
Hence, the mass for the $K^*(892)$ is shifted by
$\approx$ 10 MeV from the published value to 905 MeV.
The solutions of the Roy--Steiner equations also yield
predictions for the scattering lengths $a^{1/2}_0$ and $a^{3/2}_0$. In
contradiction to $\pi\pi$ scattering the allowed values are strongly
constrained by the data, so that there is no universal band for the
two scattering lengths. We find
$    m_\pi a^{1/2}_0 = 0.224 \pm 0.022,\quad m_\pi a^{3/2}_0 
                          = -0.0448 \pm 0.0077\nonumber$
Finally, the matching of the subthreshold parameters in the chiral and
the dispersive framework yields the following estimates for the
low--energy constants (at the scale $\mu = m_\rho$ in units of $10^{-3}$):
$   L_1^r = 1.05\pm 0.12, L_2^r = 1.32\pm 0.03, L_3^r = -4.53\pm 0.14,
   L_4^r = 0.53\pm 0.39$ and $2 L_6^r + L_8^r = 3.66\pm 1.52$.

}

\newabstract 
\label{abs:Girlanda}

\begin{center}
{\large\bf The impact of $\pi \pi$ scattering data on 
                                 SU(3) chiral dynamics}\\[0.5cm]
S.~Descotes-Genon$^1$, N.~H.~Fuchs$^2$, {\bf L. Girlanda}$^3$, and J.~Stern$^4$\\[0.3cm]
$^1$Laboratoire de Physique Th\'eorique,
91405 Orsay Cedex, France\\[0.1cm]
$^2$Department of Physics, Purdue University, West Lafayette IN 47907, USA\\[0.1cm]
$^3$European Center for Theoretical Studies in Nuclear Physics and
Related Areas, Strada delle Tarabelle 286, 38050 Trento, Italy\\[0.1cm]
$^4$Groupe de Physique Th\'eorique, Institut de Physique Nucl\'eaire, F-91406
Orsay-Cedex, France
\end{center}

Recent $\pi\pi$ scattering data from the
E865 experiment at Brookhaven \cite{e865} have confirmed the standard
scenario of large condensate in the limit of two massless flavours
\citetwo{cglprl}{noisu2}. 
The theoretical interest has therefore focused on the SU(3) chiral
dynamics.
Actually one realizes that the order parameters  appearing in the
SU(3) ChPT are experimentally observable quantities the most
directly related to chiral symmetry breakdown. The same is not
true for the two-flavour order parameters, defined with the strange
quark kept at its physical mass, which is not large enough to decouple
from the theory. In this case massive $\bar s s$ pairs, which are
abundant in the vacuum, induce e.g. an SU(2)$\times$SU(2) breaking
condensate, through OZI rule violating correlations, which adds up to
the genuine condensate. The determination of the two main
three-flavour order parameters, $\Sigma(3)$ and $F^2(3)$, respectively
the quark condensate and the pion decay constant in the $N_f=3$
chiral limit, is a delicate issue because the fluctuations of massive
$\bar s s$ pairs induce instabilities in the SU(3) chiral series. In
Ref.~\cite{noi}, a possible cure for such instabilities is explained
(see also J.~Stern's contribution to these proceedings), which amounts
to a non-perturbative 
resummation of vacuum fluctuations encoded in the low-energy constants
$L_4$ and $L_6$. A suitable class of observable is expressed in terms
of $\Sigma(3)$, $F^2(3)$ and the quark mass ratio
$r=2m_s/(m_u+m_d)$. Uncertainties due to higher chiral orders and 
theoretical constraints coming from first principles (vacuum
stability and  paramagnetic inequalities \cite{dgs}) can be suitably accounted
for in the framework of Bayesian statistical inference when fitting to
data.
It is shown that, while present $\pi\pi$ data \cite{e865} are not
accurate enough to constrain the three-flavour order parameters, they
yield a lower bound on the quark mass ratio, $r\geq 14$ at 95\%
confidence level.


\newabstract 
\label{abs:Hirenzaki}

\begin{center}
{\large\bf Formation of Meson-Nucleus Systems}\\[0.5cm]
{\bf S. Hirenzaki}\footnote{E-mail: zaki@cc.nara-wu.ac.jp} \\[0.3cm]
Department of Physics, Nara Women's University, Nara 630-8506, Japan\\[0.1cm]
\end{center}

Mesic atoms and mesic nuclei are the useful laboratory for studying the 
meson-baryon interactions and the meson properties in nuclear medium.  
Among others, pionic atoms have been long utilized for such studies, 
since they would 
provide 
valuable information on the behavior of ``real'' pions in the interior of 
the nucleus. Deeply bound pionic states were found experimentally 
for the 
first time in (d,$^3$He) 
reactions on $^{208}$Pb \citetwo{1}{2} and   
the observed spectrum 
showed an excellent agreement with our calculation made before the 
experiment \cite{3}.    
This agreement between theoretical results and data 
provides a strong confidence on the 
predictability of the theoretical model used.  Hence, 
we can investigate other exotic meson-nucleus systems using the same model. 
In this presentation, we reported recent developments of the theoretical 
studies in our group for the 
structure and formation of the meson-nucleus systems. Following 
subjects were included. 

(1) Formation of the deepest and, thus, most interesting 
1s pionic states in heavy and medium heavy nuclei  \cite{4}. Especially, we 
consider the excellent data obtained by K. Suzuki et al . \cite{5} and the 
attempts to deduce a clear evidence for partial restoration of chiral 
symmetry from the data.  

(2) Residual interaction effects in the deeply bound pionic atoms 
\cite{6}. Since the experimental data of Sn isotopes are so accurate  
($\Delta E \sim$ 20keV), the residual interaction 
effects could be important and should be evaluated. The results will be 
obtained soon \cite{6}.

(3) Structure and formation of K-mesic atoms \cite{7}. We reported several 
theoretical results for the Kaonic atom formation by (K,p) reactions 
\cite{7}.

\newabstract 
\label{abs:Nagahiro}

\begin{center}
{\large\bf
$\eta$-Nucleus interactions and in-medium properties of $N^*(1535)$ in
 chiral models
}\\[0.5cm]
{\bf H.~Nagahiro}$^1$, D.~Jido$^2$ and S.~Hirenzaki$^1$\\[0.3cm]
$^1$Department of Physics, Nara Women's University, Nara 630-8506, Japan\\[0.1cm]
$^2$European Centre for Theoretical Studies in Nuclear
Physics and Related Areas (ECT*), Villa Tambosi, Strada delle Tabarelle
286, I-38050 Villazzano (Trento), Italy
\end{center}

We investigate the properties of $\eta$-nucleus interaction and their
experimental consequences~\cite{ref:ours}. The strong coupling of
$\eta N$ to $N^*(1535)(N^*)$ makes the use of this channel particularly suited to
investigate this resonance and enables us to consider the $\eta$-mesic
nucleus as one of the doorway to investigate the in-medium properties
of $N^*$.

In this study, we investigate the $N^*$ properties in the nuclear medium
using two kinds of chiral effective models: the chiral doublet model and
the chiral unitary model.  The chiral doublet model is an extension of the SU(2)
linear sigma model for nucleon sector \citetwo{ref:18}{ref:19}. In this
model, a reduction of the mass difference of $N$ and $N^*$ in the
nuclear medium is found~\citetwo{ref:20}{ref:21}. We 
should stress that this reduction yields a curious shaped potential of
$\eta$-nucleus system. On the other hand, in the chiral unitary
model~\cite{ref:8},
$N^*$ is introduced as a resonance generated dynamically
by meson-baryon scattering. Since this
theoretical framework is quite different from the chiral doublet model, 
it is interesting to compare the
consequences of these `chiral' models for $N^*$ and $\eta$ mesic
nucleus. 

For this purpose, we calculate the (d,$^3$He) and ($\gamma$,p)
spectra for the formation
of the $\eta$-nucleus systems in the final states~\cite{ref:ours}. This (d,$^3$He) spectroscopy is
an established experimental method in the studies of the pionic bound
systems.  We conclude that
we can deduce the new information of $\eta$-nucleus interaction from these
experiments,
and by knowing the nature of the $\eta$-nucleus 
optical potential,
we will be able to study the in-medium 
properties of the $N^*$.
We believe that this research helps much the experimental activities for
the studies of the
$\eta$-nucleus systems, and the understanding of the baryon chiral
symmetries and its medium modifications.

\newabstract 
\label{abs:Suzuki}

\begin{center}
{\large\bf
  Precise measurement of deeply bound pionic 1s states of Sn nuclei
}\\[0.5cm]

{\bf K. Suzuki}$^1$ 
\\[0.3cm]
$^1$ Physik-Department, Technische Universit\"at M\"unchen,
D-85748 Garching, Germany\\[0.1cm]
\end{center}

A goal of this work is to measure the degree of chiral symmetry
restoration in a nuclear medium through the determination of
the isovector $\pi N$ interaction parameter in the pion-nucleus
potential by studying deeply bound 1s states of $\pi^-$ in
heavy $N>Z$ nuclei. 
We performed a systematic experimental studies of 1s $\pi^-$
states in a series of Sn isotopes, which were produced with the
Sn($d$,$^3$He) reactions.
One of the advantages of using Sn isotopes is that we can produce the
1s $\pi^-$ states as the most dominant quasi-substitutional states,
$(1s)_{\pi^-}(3s)_n^{-1}$, because of the presence of the 3s orbital
near the Fermi surface, as theoretically predicted~\cite{Umemoto}.
Another merit is to make use of isotopes over a wide range of
$(N-Z)/A$ to test the isospin dependence~\cite{Kienle:01}.

We observed spectra, $d^2\sigma/(dEd\Omega)$, on mylar-covered
$^{116}$Sn, $^{120}$Sn, $^{124}$Sn targets as function of the $^3$He
kinetic energy~\cite{ksuzuki}, 
from which we determined precisely the 1s binding energies
($B_{1s}$) and widths ($\varGamma_{1s}$) as summarized in the
table~\ref{tab:BE} below.
\begin{table*}[htb]
\begin{center}
\label{tab:BE}
\begin{tabular}{ccccc}
\hline
Isotope & $B_{1s}~{\rm [MeV]}$         & $\Delta B_{1s}~{\rm [MeV]}$ 
        & $\varGamma_{1s}~{\rm [MeV]}$ & $\Delta \varGamma_{1s}~{\rm [MeV]}$\\ 
\hline
$^{115}$Sn & 3.906 & $\pm 0.024$ & 0.441 & $\pm 0.087$\\
$^{119}$Sn & 3.820 & $\pm 0.018$ & 0.326 & $\pm 0.080$\\
$^{123}$Sn & 3.744 & $\pm 0.018$ & 0.341 & $\pm 0.072$\\
\hline
\end{tabular}
\end{center}
\end{table*}

The obtained data of binding energies and widths of the 1s $\pi^-$
states in $^{115,119,123}$Sn combined with those of symmetric light
nuclei ($^{16}$O, $^{20}$Ne and $^{28}$Si) yielded
$b_1 = -0.116 \pm 0.007~m_{\pi}^{-1}$~\cite{ksuzuki}.
The error includes both statistical and systematic errors. One of the
main sources of the systematic error originates in uncertainty of
neutron density distributions.
The new, yet unpublished, pp-scattering data from RCNP/Osaka~\cite{RCNP}
are in complete agreement with the antiprotonic atom data we use.

The magnitude of the $|b_1|$ is significantly enhanced over
the free $\pi N$ value
, which translates into a
reduction of ${f_\pi }^2$~\cite{Kolomeitsev:02} as
$
R  = \frac{b_1^{\rm free}}{b_1} =0.78 \pm 0.05\nonumber
     \approx \frac{b_1^{\rm free}}{b_1^* (\rho_e)} 
     \approx \frac{f_{\pi}^* (\rho_e)^2}{f_{\pi}^2}
     \approx 1 - \alpha \rho_e,$
with an effective density $\rho_e\approx 0.6 \rho_0$.
This hence implies that the chiral order parameter,
$f_{\pi}(\rho)^2$, would be reduced by a factor of $\approx 0.64$
at the normal nuclear density $\rho = \rho_0$.


\newabstract 
\label{abs:Cargnelli}

\begin{center}
{\large\bf DEAR - Kaonic Hydrogen: First Results}\\[0.5cm]
G.~Beer$^i$, A.M.~Bragadireanu$^{a,e}$, {\bf M.~Cargnelli$^d$,}
C.~Curceanu (Petrascu)$^{a,e}$, J.-P.~Egger$^{b,c}$,
H.~Fuhrmann$^d$, C.~Guaraldo$^a$, M.~Iliescu$^a$,
T.~Ishiwatari$^d$, K.~Itahashi$^g$, M.~Iwasaki$^f$, P.~Kienle$^d$,
B.~Lauss$^h$, V.~Lucherini$^a$, L.~Ludhova$^b$, J.~Marton$^d$,
F.~Mulhauser$^b$, T.~Ponta$^{a,e}$, L.A.~Schaller$^b$,
R.~Seki$^{j,k}$, D.~Sirghi$^a$, F.~Sirghi$^a$,
P.~Strasser$^f$ , J.~Zmeskal$^d$ \\[0.3cm]

$^a$INFN - Laboratori Nazionali di Frascati; $^b$Universite de
Fribourg; $^c$Universite de Neuch$\hat{a}$tel; $^d$Institute for
Medium Energy Physics, Vienna; $^e$Institute of Physics and
Nuclear Engineering, Bucharest; $^f$RIKEN, Saitama; $^g$Tokyo
Institute of Technology; $^h$University of California and
Berkeley; $^i$ University of Victoria; $^j$California Institute of
Technology;
$^k$California State University \\[0.1cm]
\end{center}

The DEAR\footnote{DA$\Phi$NE Exotic Atom Research, conducted at
the Frascati electron positron collider} experiment \cite{rf:1}
measures the energy of X-rays emitted in the transitions to the
ground states of kaonic hydrogen. The shift $\epsilon$ and the
width $\Gamma$ of the 1s state are related to the real and
imaginary parts of the complex S-wave scattering length by the
Deser Trueman formula. \

\begin{figure} [h]
 \centering
 \begin{minipage}[c]{.45\textwidth}
   \centering
   \caption{Background subtracted energy spectrum of kaonic hydrogen X ray transitions.}
   \label{fig:1}
 \end{minipage}
 \hfill
 \begin{minipage}[c]{.45\textwidth}
   \centering
   \includegraphics [angle=-90,scale=0.19]{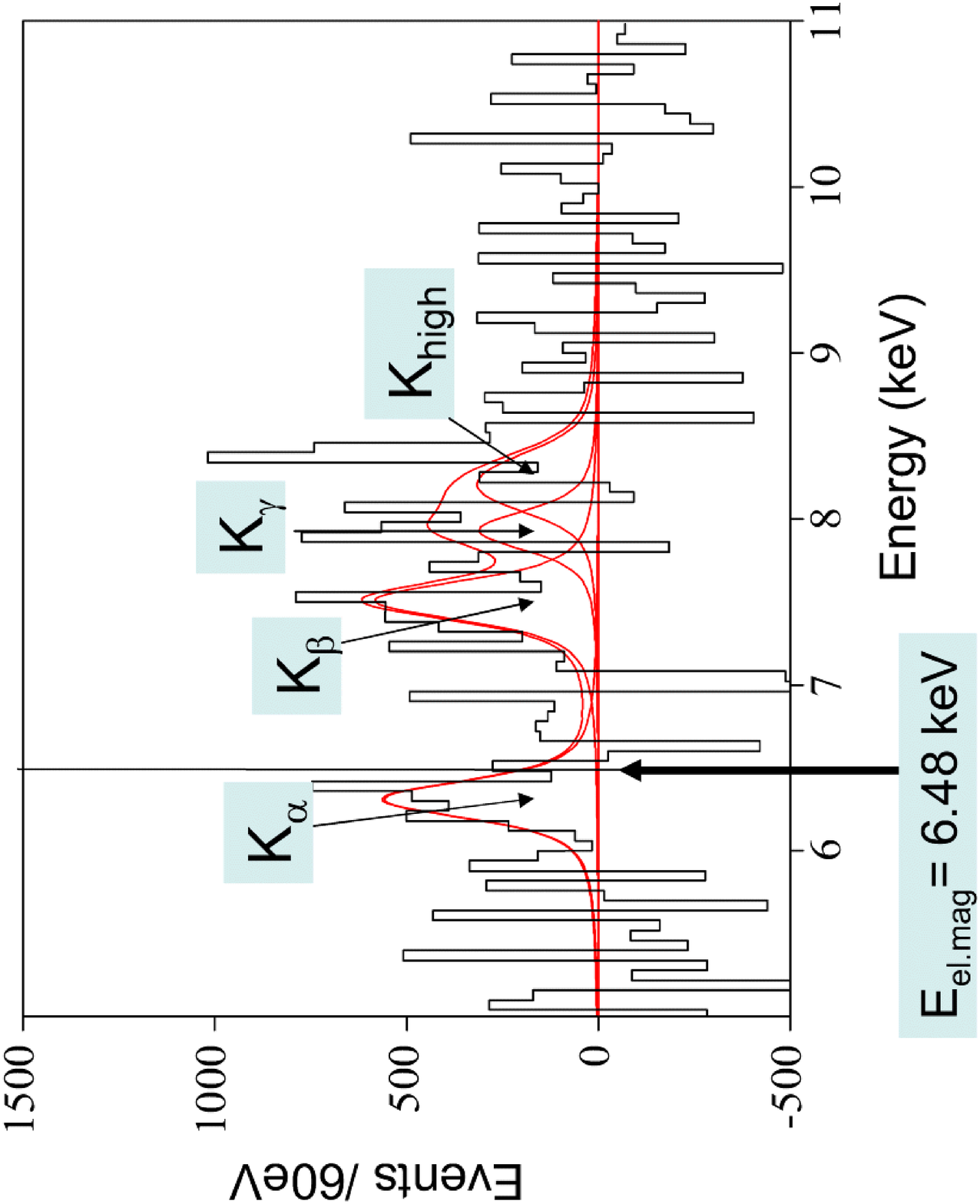}
 \end{minipage}
\end{figure}

The preliminary results are: $\epsilon$= - 202 $\pm$ 45 eV  and
$\Gamma$= 250 $\pm$ 138 eV. Both values are smaller then those
from the previous experiment \cite{rf:2} and consistent with
recent theoretical studies \cite{rf:3}. 

\bigskip

Part of the work was supported by "Transnational access to
Research Infrastructure" (TARI) Contract No. HPRI-CT-1999-00088.

\newabstract 
\label{abs:Kolomeitsev}

\begin{figure}
\centerline{\parbox{5cm}{
\includegraphics[width=5cm,clip=true]{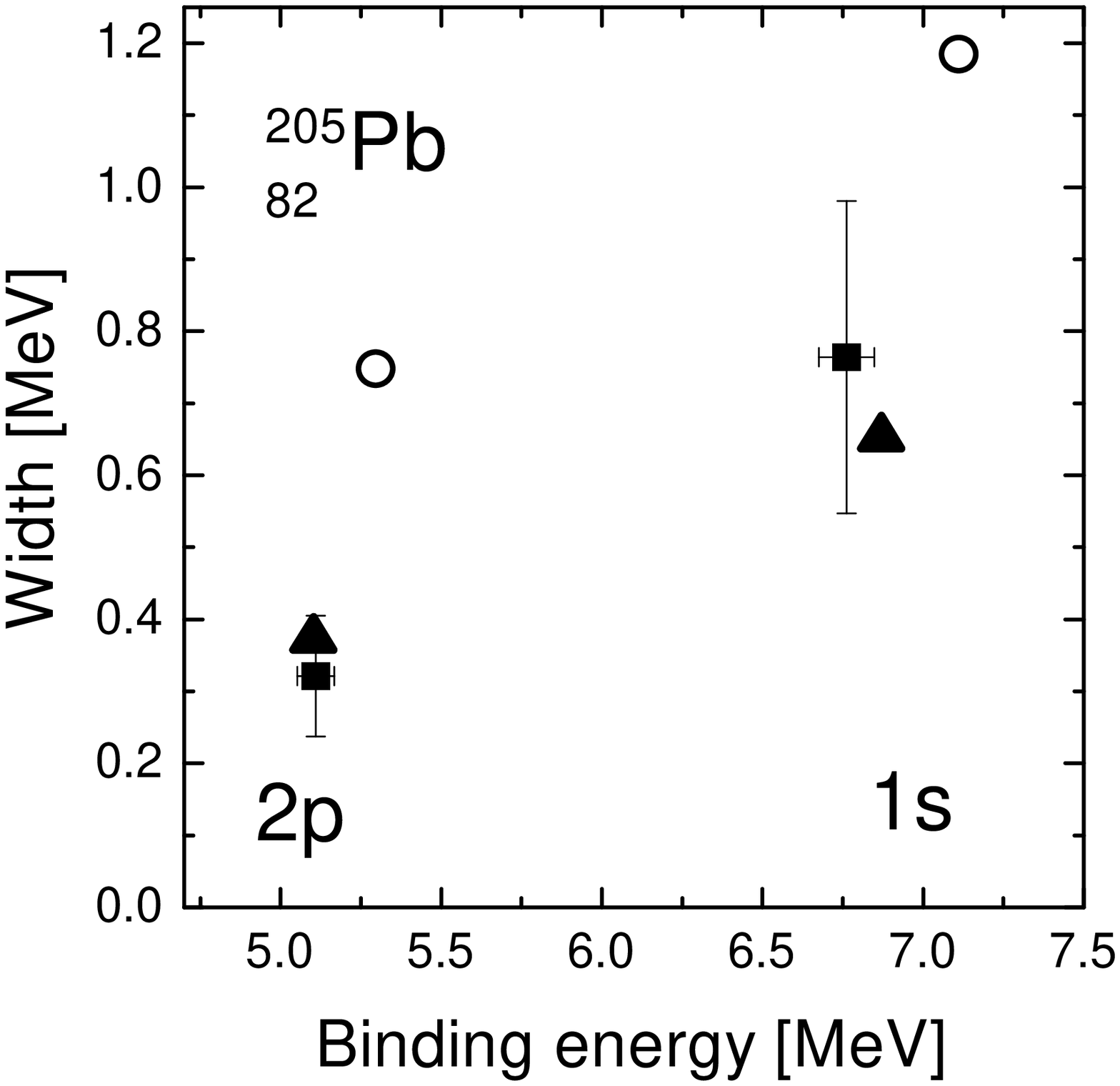}}\qquad
\parbox{5cm}{\includegraphics[width=5cm,clip=true]{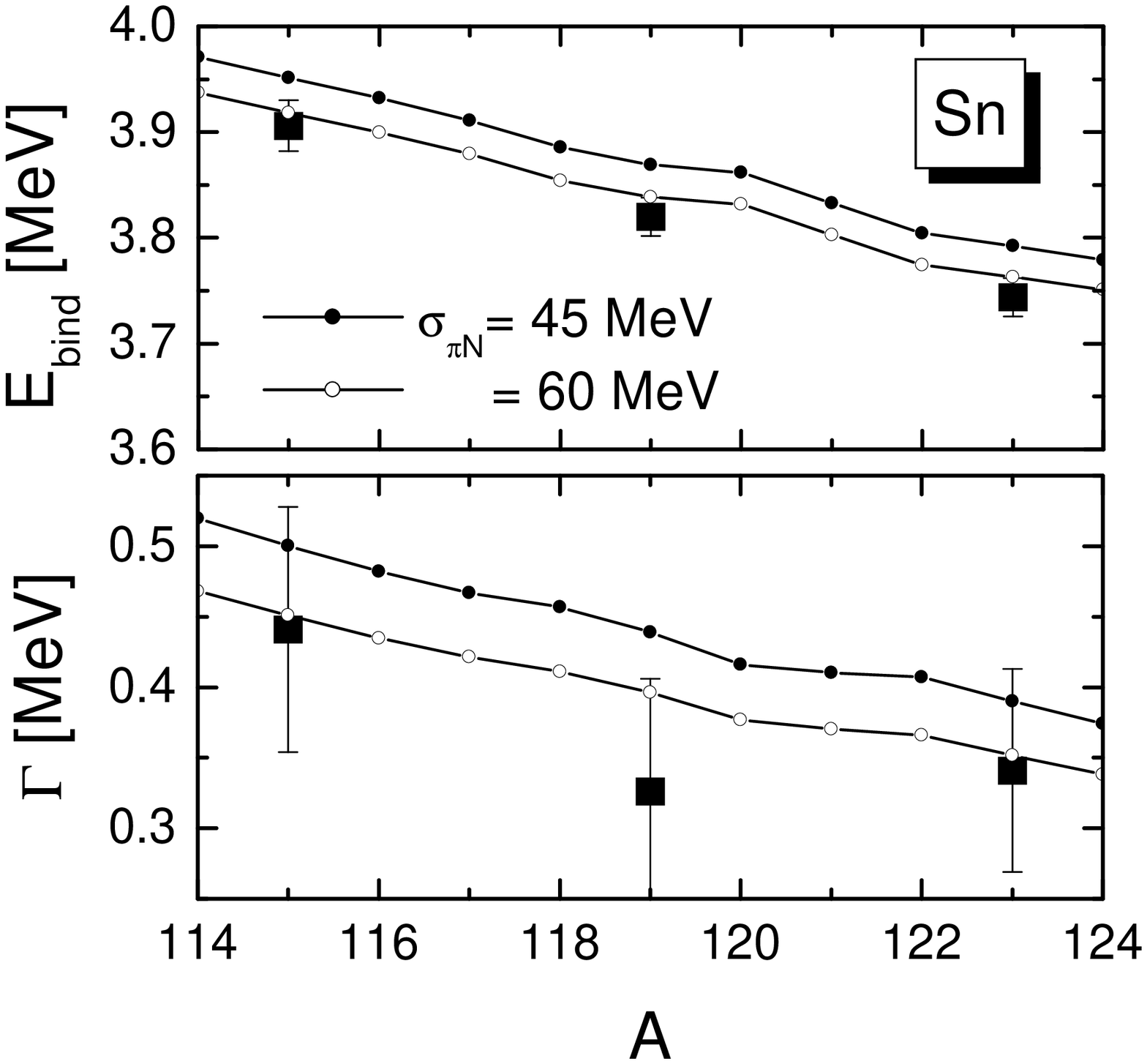}}}
\caption{
Left panel:
Binding energies and widths of pionic $1s$ and $2p$ states
in  $^{205}$Pb isotopes for the energy-dependent (full triangles) and
energy-independent (open circles) s-wave polarization  operator.
Right panel:
Curves show binding energies and widths of pionic $1s$ states in Sn isotopes.
for different values of the $\pi N$ sigma term. Data are from
\cite{data}
}\label{fig}
\end{figure}

\begin{center}
{\large\bf Chiral dynamics and  pionic states of Pb and Sn isotopes}\\[0.5cm]
{\bf E.E.~Kolomeitsev}$^1$, N.~Kaiser$^2$, and W.~Weise$^{2,3}$\\[0.3cm]
$^1$NBI,  Copenhagen,
 $^2$TU M\"unchen, Garching,
 $^3$ECT*, Trento
\end{center}
In recent papers~\citetwo{kkw}{kw} we have re-investigated the issue
of missing repulsion in pionic atoms
from the point of view of the distinct explicit energy dependence
of the pion-nuclear polarization operator.
The starting point is the energy- and  momentum-dependent
polarization operator $\Pi(\omega,\vec
q;\rho_p,\rho_n)$\,. In the limit of very low proton and neutron
densities, $\rho_{p,n}$, the pion self-energy reduces to
$\Pi=-(T^+\,\rho+T^-\, \delta\rho)$ with $\rho=\rho_p+\rho_n$ and
$\delta \rho=\rho_p-\rho_n$, where $T^{\pm}$ are the isospin-even
and isospin-odd off-shell $\pi N$ amplitudes.
Terms of  sub-leading orders in density are incorporated within in-medium chiral effective field theory,
following~\cite{kw}. Double scattering
corrections are fully incorporated at 2-loop order. Absorption
effects and corresponding dispersive corrections appear  at the
3-loop level and through  short-distance dynamics
parameterized by contact terms, not explicitly calculable
within the effective low-energy theory. The imaginary parts
associated with these terms are well constrained  by the
systematics  of observed  widths of pionic atom levels throughout
the periodic table. The real part of the s-wave absorption term ($\Re
B_0$)  is still the primary
source of theoretical uncertainty. As suggested also by the detailed
analysis of the pion-deuteron scattering length
we utilize $\Re B_0=0$.  The canonical
parameterization  of p-wave parts is included as well.
We solve the
Klein-Gordon equation (KGE)
$[(\omega-V_c)^2+\vec\nabla^2-m_\pi^2-
\Pi(\omega-V_c;\rho)]\phi=0 $ in the local density approximation.
The explicit energy dependence of $\Pi$ requires that
the Coulomb potential $V_c$ must be introduced in the
canonical gauge-invariant way.
With input specified in details in~\cite{kkw}, we have solved
KGE with the explicitly energy dependent pion
self-energy.
The results for the binding energies
and widths of $1s$ and $2p$ states in pionic $^{205}$Pb are shown
in Fig.~\ref{fig} in comparison with the
outcome of "standard" phenomenological calculations using a
energy independent s-wave optical potential (empty circles).
The explicit energy  dependence in $T^{\pm}$ results in
considerably batter agreement with experimental data.
The replacement
$\omega \to \omega-V_c>m_\pi$ increases the repulsion in $T^-$
and disbalances the "accidental" cancellation between the $\pi N$
sigma term $\sigma_{N}$ and the range term proportional to
$\omega^2$ in $T^+$, such that $T^+(\omega-V_c)<0$ (repulsive).
The combined
effect from uncertainties in $\Re B_0$, in the radius and shape of the
neutron density distribution falls within the experimental errors.
Using the same scheme we have
predicted binding energies and widths for pionic $1s$ states bound
to a  chain of Sn isotopes.  Results are shown in Fig.~\ref{fig} in
comparison with the   experimental data~\cite{data}.
This figure also gives an
impression of the sensitivity with respect to variations of the $\pi N$ sigma term.

\clearpage

\newabstract 
\label{abs:Friedman}

\begin{center}
{\large\bf Chiral restoration from pionic atoms}\\[0.5cm]
{\bf E. Friedman}~$^1$ 
and A. Gal \footnote{Supported by the
Israel Science Foundation grant No. 131/01.} \\[0.3cm]
$^1$Racah Institute of Physics, The Hebrew University, Jerusalem 91904,
Israel\\[0.1cm]
\end{center}

A major component of the so-called anomalous s-wave repulsion
in pionic atoms, as extracted from fits of optical potentials
to pionic atom data, is due to the isovector s-wave $\pi N$
amplitude $b_1$ which comes out too repulsive compared to the free
$\pi N$ amplitude [1].
Average values of $b_1$ with added uncertainties due to
neutron distributions are summarized in the table, for the standard
Ericson-Ericson model, including double-scattering 
and angle-transformation terms. 
The free pion-nucleon value 
is $b_1^{\rm free}=-0.0885^{+0.0010}_{-0.0021}~m_\pi ^{-1}$.
`Deep' refers to deeply bound 1$s$ states in $^{115,119,123}$Sn and
$^{205}$Pb.  
\begin{center}
\begin{tabular}{lccccc}
\hline
data & `global 2'& `global 3'& light $N=Z$ & light $N=Z$ & `deep'\\
 &$^{12}$C to $^{238}$U&$^{20}$Ne to $^{238}$U& + light $N>Z$&
+`deep' & \\
 & & & 1$s$ only&1$s$ only & 1$s$ only\\
\hline
points       & 120 & 100 & 22 & 20 & 8\\
$b_1 (m_\pi ^{-1})$ &$-$0.108$\pm$0.007 &$-$0.104$\pm$0.006&
$-$0.099$\pm$0.014 &$-$0.104$\pm$0.013 & $-$0.130$\pm$0.036 \\
\hline
\end{tabular}
\end{center}
It is obvious that only {\it large} data sets provide conclusive
evidence for the modification of $b_1$ in the nuclear medium.

The in-medium $s$-wave interaction of pions has been discussed
recently by Weise \cite{Wei01}
in terms of partial restoration of chiral symmetry, leading to
a density-dependent $b_1(\rho)$ which essentially removes the anomaly
\cite{Fri02}. Alternatively, Kolomeitsev et al. \cite{KKW03} imposed the
minimal substituion $E \to E - V_{c}$ using energy-dependent chirally
expanded amplitudes for {\bf q}=0. 
We have found [5] this approach
to disagree with the data, for large data sets, and with the value
of the free $\pi N$ amplitude $b_1$, whereas good agreement is 
obtained using the
empirical on-shell energy dependence as suggested long ago by Ericson
and Tauscher [6].

\newabstract 
\label{abs:Rusetsky}

\begin{center}
{\large\bf Introduction to the ChPT for Heavy Nuclei}\\[0.5cm]
L. Girlanda$^1$, {\bf A. Rusetsky}$^{2,3}$, and W. Weise$^{1,4}$\\[0.3cm]
$^1$ECT*, Strada delle Tabarelle 286, I-38050 Villazzano (Trento), Italy\\[0.1cm]
$^2$HISKP (Theorie), University of Bonn, Nu\ss{}allee 14-16, 53115 Bonn, Germany\\[0.1cm]
$^3$HEPI, Tbilisi State University, University St.~9, 380086 Tbilisi, Georgia\\[0.1cm]
$^4$Physik-Department, Technische Universit\"at M\"{u}nchen,
D-85747 Garching, Germany
\end{center}

Recent accurate measurements of deeply bound (1$s$) states 
of pionic atoms formed with $Pb$ and $Sn$ isotopes~\cite{Suzuki:2002ae} have 
triggered renewed interest in the underlying mechanisms governing $S$-wave 
pion-nucleus interactions, raising the quest 
for  ``fingerprints of chiral restoration''~\cite{Weise:sg}. Latest 
theoretical investigations~\cite{KW} based on ChPT, have focused on the 
calculation of the in-medium shift of the pion mass.
Alternatively, the energy-dependent pion self-energy operator
has been extrapolated, using ChPT input and
the local density approximation, to calculate directly the pionic $1s$ and
$2p$ level shifts and widths~\cite{Kolo}.
In order to have the reliable predictions, however,
all these analyzes should be carefully reexamined in a
ChPT framework, where the nucleus is consistently treated as a finite
system.

In our paper~\cite{preparation} we give a systematic
formulation of ChPT in a non-uniform fermionic background,
which corresponds to the (finite) nucleus. The construction of the 
relativistic field-theoretical counterpart of the multiple-scattering 
theory on the static nuclear target is described in detail. It is shown,
that in this framework
chiral symmetry plays a crucial role, leading to the suppression of the
non-local effects in the nuclear matrix elements.

Using the formulated framework, we present a systematic derivation of the
pion-nucleus optical potential at $O(p^5)$ in ChPT, including a full pattern 
of electromagnetic and strong isospin-breaking effects. In the future, we plan 
to perform similar calculations at $O(p^6)$ -- beyond the linear density
approximation for the optical potential.

\newabstract 
\label{abs:Wirzba}

\begin{center}
{\large\bf  ChPT in the nuclear medium - the generating functional 
approach}\\[0.5cm]
{A. Wirzba}\\[0.3cm]
HISKP (Theorie), Universit\"at Bonn, Nussallee 14-16, D-53115 Bonn, 
Germany
\end{center}

We report on  a systematical study of the properties of pions in nuclear matter
under the presence of external sources~\cite{Meissner:2001gz}.  For this
purpose the methodology of the generating functional formalism of ChPT is
applied which originates back to Ref.~\cite{Gasser:1983yg} 
and which has been transcribed to nuclear
matter by Oller~\cite{Oller:2001sn}.  The corresponding in-medium lagrangian
is non-covariant as well as non-local and involves solely pions and external
sources.

Within this approach the derived (chiral) power counting rules separate for
the standard and the non-standard scenario where either the residual nucleon
energies are of the order of the pion mass or of the nucleonic kinetic energy,
respectively. The scales of applicability of these two perturbative expansions
are established as $\sqrt{6 \pi} f_\pi \simeq 0.7\,{\rm GeV}$ and ${3 \pi^2
  f^2}/{M_N} \simeq 0.27\,{\rm GeV}$, respectively, instead of the vacuum
scale $4\pi f_\pi\simeq 1.2\,{\rm GeV}$.

We present a systematic analysis of the in-medium contributions to the quark
condensates, pion propagators, pion masses, and pion couplings to the
axial-vector, vector and pseudoscalar currents for symmetric and non-symmetric
nuclear matter.  In particular, it is found that the chiral symmetry-breaking
contributions to the quark condensate are subleading, and that the
upward-shift of the in-medium mass of the $\pi^-$ can be traced back to the
Weinberg term which, however, is amplified by a factor of two by a
corresponding reduction in the wave-function renormalization coefficient.

The result of Ref.\,\cite{Thorsson:1995rj} (see also
\citetwo{Wirzba:1995sh}{Kirchbach:1996xy}) is confirmed that the vacuum pion decay
constant splits in the medium into a time-like and space-like part, and that
is is the time-like one that enters in the in-medium versions of the
Gell-Mann--Oakes--Renner and PCAC relations.
  
Finally, we report that the in-medium pion-pion scattering at leading order is
${\cal O}(q^3)$ and therefore belongs to the non-standard counting scenario.
The dominate process at low three-momentum and density is the twice-iterated
Weinberg-Tomozawa scattering off the nucleon which is of S-wave nature and
repulsive.
 
A.W. thanks his coauthors Ulf-G.~Mei{\ss}ner and Jos\'e~A.~Oller of
Ref.\,\cite{Meissner:2001gz} and acknowledges support of the Forschungszentrum
J\"ulich GmbH under Contract 41445400 (COSY-067).

\newabstract 
\label{abs:Petrascu}

\begin{center}
{\large\bf Future precision measurements on kaonic
 hydrogen and kaonic deuterium with SIDDHARTA}\\[0.5cm]
{\bf C. Curceanu (Petrascu)} (for the SIDDHARTA Collaboration)\\[0.4cm]
INFN - Laboratori Nazionali di Frascati, C.P. 13, Via E. Fermi 40,
I-00044 Frascati, Italy \\[0.1cm]

\end{center}

 After the most precise measurement on kaonic hydrogen performed up
 to now by DEAR at the DA${\Phi}$NE collider of Laboratori Nazionali di
 Frascati in 2002, result  presented at this meeting \cite{Cargnelli},
 a new precision measurement of kaonic hydrogen and kaonic
 deuterium is envisaged in the framework of the new SIDDHARTA
 project.   
 
 The objective of 
 SIDDHARTA (SIlicon Drift Detector for Hadronic Atom Research
 by Timing Application) is the precision measurement of the K$_{\alpha}$
 line shift and width, due to the strong interaction,  in  kaonic
 hydrogen and a similar measurement - the first one - in kaonic deuterium.
 The aim is a precision determination of the antikaon-nucleon isospin
 dependent scattering lengths.
 An accurate determination of the K$^-$N isospin dependent
scattering lengths will place strong constraints on the low-energy K$^-$N
dynamics, which in turn constraints the SU(3) description of chiral symmetry
breaking \cite{Guaraldo}. Crucial information about the nature of chiral symmetry breaking, and
to what extent the chiral symmetry must be broken, is provided by the
calculation of the meson-nucleon sigma terms \cite{Reya}.

 SIDDHARTA represents the natural continuation of DEAR taking
 advantage, however, of a completely newly designed setup which
 guarantees a near 4$\pi$ solid angle coverage, and, as detector,
 silicon drift detectors (SDD) which exhibit the same performance
 of CCDs as far as efficiency and energy resolution is concerned,
 but can be triggered, thus allowing an increasing of
 the signal/background ratio of orders of magnitude.
 
 The trigger is given by the entrance of the charged kaon in the target volume.
The event can be identified and measured with high accuracy and low
contamination by the use of a three-scintillator telescope, synchronized with
the bunch frequency.
First tests of a prototype SDD array, performed at the Beam Test 
Facility (BTF) of LNF, gave excellent results.
                                                     
 In addition to the precision kaonic hydrogen and deuterium measurements, in
 SIDDHARTA it is planned a measurement of kaonic helium and a feasibility
 study of sigmonic hydrogen.
 
 \vspace{.4cm}
 We thank  the DA${\Phi}NE$
  BTF  group for the very good cooperation and team
 work.
 
 Part of the work was supported by a  ``Transnational Access to Research
Infrastructure" (TARI) Contract  No.  HPRI-CT-1999-00088.

\newabstract 
\label{abs:Krewald}

\begin{center}
{\large\bf 
Lifetime of Kaonium
}\\[0.5cm]
{\bf S. Krewald}$^1$, R. H. Lemmer$^2$, and F. P.  Sassen$^1$\\[0.3cm]
$^1$Institut f\"ur Kernphysik, Forschungszentrum, 52425 J\"ulich, Germany\\[0.1cm]
$^2$Nuclear and Particle Theory Group, University of Witwatersrand, Johannesburg, 
Private Bag 3, WITS 2050, South Africa\\[0.1cm]
\end{center}

The scalar mesons $f_0(980)$ and $a_0(980)$ have been interpreted as two quark-antiquark ($q^2 \bar{q}^2$)
states by Jaffe and by Achasov,
 whereas Weinstein and Isgur suggest a quasibound Kaon-Antikaon ($K\bar{K}$) structure.
Here we present  a prediction for the lifetime of the hadronic atom kaonium based on the
mixing with scalar mesons having a ($K\bar{K}$) structure. 

The small binding energy of
about 20 MeV of the $K\bar{K}$ meson suggests a non-relativistic 
 effective field theory approach\cite{GLRG01}.
 With this in mind we  use the standard $SU(3)_V\times SU(3)_A$
 Lagrangian to describe the dynamics of the
 $K\bar K$ interaction\cite{Jansen95} and decay via the exchange of
 $\rho,\omega,\phi,K^*....$ vector mesons, assuming
 $SU(3)$ symmetry for the coupling constants.
  Then we replace the 
 OBE potentials by phase--equivalent potentials of the Bargmann
 type\cite{Bargmann49} that give  rise to the same scattering length and
 effective range. For these potentials,
 the associated Jost functions can be constructed explicitly 
so that both the scattering and bound state properties of the
 $K\bar K$ system are determined without further approximation.

The hadronic atom kaonium so far has found little attention in the literature\cite{SWycech}.
We construct the Jost function of kaonium and determine the bound states by the
Kudryavtsev--Popov equation\cite{KP79},
Full details can be found in ref.\cite{KLB}.
Below, we present the prediction for the lifetime of kaonium based on the assumption of a  
$K\bar{K}$ meson structure. The full J\"ulich meson exchange model contains a hard component 
in addition to the $K\bar{K}$ meson structure. This reduces the lifetime by a factor 3.

\begin{center}
 \begin{tabular}{cccc}
Level&$\lambda$&$\Delta E-i\Gamma/2$ (keV)&Lifetime ($\times 10^{-18}$ sec)\\
{$3^{rd}$}&$0.2491+0.0005i$&$0.003-0.002i$& $199$ \\ 
{$2^{nd}$}&$0.3318+0.0009i$&$0.007-0.004i$&$84$ \\
{$1^{st}$}&$0.4965+0.0020i$&$0.023-0.013i$&$25$ \\
{Ground}&$0.9863+0.0079i$&$0.180-0.103i$&$3.2$ \\ 
	\end{tabular} 
\end{center}

\newabstract 
\label{abs:Jido}

\begin{center}
{\large\bf Chiral dynamics of the two $\Lambda(1405)$ states}\\[0.5cm]
{\bf D. Jido}$^1$,  J.A. Oller$^2$, E. Oset$^3$, A. Ramos$^4$ and 
U.-G. Mei\ss{}ner$^5$\\[0.3cm]
$^1$ECT$^{*}$, Villa Tambosi, Strada delle Tabarelle 286, I-38050 
Villazzano (Trento), Italy\\[0.1cm]
$^2$Departamento de F\'{\i}sica, Universidad de Murcia, 30071 
Murcia, Spain\\[0.1cm]
$^3$Departamento de F\'{\i}sica Te\'orica and IFIC,
Centro Mixto Universidad de Valencia-CSIC, Institutos de
Investigaci\'on de Paterna, Aptd. 22085, 46071 Valencia, Spain\\[0.1cm]
$^4$Departament d'Estructura i Constituents de la Mat\`eria,
Universitat de Barcelona, Diagonal 647, 08028 Barcelona, Spain\\[0.1cm]
$^5$ HISKP (Theorie), University of Bonn, Nu{\ss}alle 14-16, D-53115 Bonn, Germany
\end{center}

The $\Lambda(1405)$ has been a long-standing example of a
dynamically generated resonance appearing naturally in scattering theory with
coupled meson-baryon channels with strangeness $S=-1$. Modern chiral
formulations of the meson-baryon interaction within unitary frameworks all lead to
the generation of this resonance, which is seen as a near Breit-Wigner form in
the mass distribution of $\pi \Sigma$ states with isospin $I=0$ in hadronic 
production processes.

As a theoretical consequence, studying the analytical structure of the
scattering amplitude in the $S=-1$, $I=0$ channel obtained by the chiral
unitary approach, we find two poles very close to the $\Lambda(1405)$
\cite{Jido:2003cb}. Both of them are sitting between the $\pi\Sigma$ and $\bar
KN$ thresholds, and these two resonances are quite close
but different. The important nature of them is that the one at lower energies has
a larger width and a stronger coupling to the $\pi \Sigma$ states, while the
other at higher energies couples mostly to the $\bar{K}N$ states and has a
smaller width.  This is the main finding of the present work, thus we conclude
that there is not just one single $\Lambda(1405)$ resonance, but {\em two}, and
that what one sees in experiments is a {\em superposition} of these two states. 

The existence of the two pole is strongly related to the flavor 
symmetry. The underlying $SU(3)$ structure of the chiral Lagrangians
implies that, together with the pole around the $\Lambda(1670)$, a singlet 
and two octets of dynamically generated resonance should appear, but the
dynamics of the problem makes the two octets degenerate in the case of exact
$SU(3)$ symmetry.  In the physical limit, the breaking of the $SU(3)$ symmetry
resolves the degeneracy of the octets and two distinct octets appear.
The breaking of the degeneracy has as a consequence that one of the $I=0$
octet poles appears quite close to the singlet pole.

The important theoretical finding here should
stimulate new experiments exciting the $\Lambda(1405)$ resonance.
We suggest that it is possible to
find out the existence of the two resonances by performing different experiments,
since in different experiments the weights by which the two resonances are
excited are different.  In this respect we call the attention to one reaction, 
$K^-p \to \Lambda(1405) \gamma$, which gives much weight to the resonance which
couples strongly to the $\bar{K}N$ states and, hence, leads to a peak structure
in the invariant mass distributions which is narrower and appears at higher
energies.

\newabstract 
\label{abs:Lyubovitskij}

\begin{center}
{\large\bf Chiral dynamics of baryons as bound states of  
constituent quarks}\\[0.5cm]
V.\ E.\ Lyubovitskij, Amand \ Faessler, Th.\ Gutsche, 
K. \ Pumsa-ard, P. \ Wang \\[0.3cm]
Institut f\"ur Theoretische Physik, Universit\"at T\"ubingen,\\
Auf der Morgenstelle 14, D-72076 T\"ubingen, Germany 
\end{center}
We developed a manifestly Lorentz invariant chiral quark model for the 
study of baryons as bound states of constituent quarks in an extention 
of our previous approach~\citetwo{Ref1}{Ref2}.  
Our investigations are related to the ongoing and planned experiments 
at ELSA, JLab, MAMI, MIT, NIKHEF 
on improved measurements of the elastic baryon form factors and 
their decay characteristics. The approach is based on a non-linear 
chirally symmetric Lagrangian, which involves effective 
degrees of freedom - constituent quarks and the chiral (meson) fields. 
In the first step, this Lagrangian can be used to perform a dressing of 
the constituent quarks by a cloud of light pseudoscalar mesons and other 
heavy states using the calculational technique developed by Becher and 
Leutwyler~\cite{Ref3} which is based on the infrared regularization of 
loop integrals. We calculate the dressed transition operators 
with a proper chiral expansion which are relevant for the interaction of 
quarks with external fields in the presence of a virtual meson cloud. 
Next, these operators are used in the calculation of baryon matrix elements 
using an effective Lagrangian describing the coupling of a baryon field to 
the interpolating three-quark current~\cite{Ref2}. In Table I   
we summarized our results for canonical properties of nucleon and $\Delta$: 
nucleon magnetic moments and charge radii,  the helicity amplitudes $A_{1/2}$ 
and $A_{3/2}$ of the $\Delta \to N \gamma$ transition and the ratio $E2/M1$. 
For comparison we present the experimental data (central values).

\vspace*{.25cm} 

{\bf Table I.} Static characteristics of nucleon and $\Delta$
\begin{center}
\begin{tabular}{|l|l|l|l|l|l|l|l|l|l|}
\hline\hline
       &  $\mu_p$ & $\mu_n$ & $r^E_p$ &  $<r^2>^E_n$   
                            & $r^M_p$ &  $r^M_n$ 
                            & $A_{1/2}$ 
                            & $A_{3/2}$ 
                            & $E2/M1$ \\
       &          &         &         (fm) & (fm$^2$)
                            &         (fm) & (fm)
                            & (GeV$^{-1/2}$) 
                            & (GeV$^{-1/2}$) &  (in \%)\\
\hline
Model   &  2.79    &  -1.91 & 0.85 &  -0.12   
                            & 0.84 &  0.85 
                            & - 0.132
                            & - 0.253
                            & - 2.5 \\
\hline
Exp    &  2.79    &  -1.91  & 0.86 
                            & -0.116 
                            & 0.86 
                            & 0.88 
                            & - 0.135 
                            & - 0.255 
                            & - 2.5  \\
\hline\hline
\end{tabular}
\end{center}

\vspace*{.3cm}

This work was supported by the Deutsche Forschungsgemeinschaft (DFG) 
under Contract Nos. FA67/25-3 and GRK683. V.E.L. thanks the ECT* 
for partial support during the meeting "Hadronic Atoms ".

\newabstract 
\label{abs:Dax}

\begin{center}

{\large\bf Towards the most precise test of bound state QED}\\[0.5cm]

{\small{
W.~Amir$^{1}$,
A.~Antognini$^{4}$,
F.~Biraben$^{1}$,
M.~Boucher$^{3}$,
C.A.N.~Conde$^{2}$,
{\bf A.~Dax}$\mathbf{^{6,7}}$,
S.~Dhawan$^{6}$,
L.M.P.~Fernandez$^{2}$,
T.W.~H\"ansch$^{4}$,
F.J.~Hartmann$^{5}$,
V.-W.~Hughes$^\dagger$$^{6}$,
O.~Huot$^{3}$,
P.~Indelicato$^{1}$,
L.~Julien$^{1}$,
S.~Kazamias$^{1}$,
P.~Knowles$^{3}$,
F.~Kottmann$^{8}$,
Y.-W.~Liu$^{10}$,
J.~Lopes$^{2}$,
L.~Ludhova$^{3}$,
C.~Monteiro$^{2}$,
F.~Mulhauser$^{3}$,
F.~Nez$^{1}$,
R.~Pohl$^{4,7}$,
P.~Rabinowitz$^{9}$,
J.M.F.~dos~Santos$^{2}$,
L.~Schaller$^{3}$,
J.-T.~Shy$^{10}$,
D.~Taqqu$^{7}$,
J.F.C.A.~Veloso$^{2}$\\[0.3cm]

Collaboration R-98-03.1: PARIS$^1$ -- COIMBRA$^2$ -- FRIBOURG$^3$ 
-- MPQ MUNICH$^4$ -- 
TU MUNICH$^5$ -- YALE$^6$ -- PSI$^7$ -- ETHZ$^8$ -- PRINCETON$^9$
-- HSINCHU TAIWAN$^{10}$}}
\end{center}
\vspace{-1ex}
\normalsize
An experiment is currently going on at Paul Scherrer Institute to determine the Lamb shift 2S$\rightarrow$2P in muonic hydrogen.
From this result the rms proton charge radius can be deduced with 10$^{-3}$ accuracy a factor of 30 better than presently
known \cite{Kot98}.\\
The principal motivation for this experiment comes from improved determinations of the Lamb shift in hydrogen. At present the
interpretation of the 1S-Lamb shift in hydrogen is limited by the uncertainty in the proton charge radius. This limit can be improved
an order of magnitude via combining a
Lamb shift measurement in $\mu$p with that in hydrogen.\\
A second motivation comes from the variety of available electron scattering data and the still going on discussion of how to derive
the rms proton charge radius from them. Since Hof\-statters pioniering experiment in 1963 the given value of the proton charge radius
changed more and more from 0.805(11)fm \cite{Han63} to nowadays 0.895(18)fm \cite{Sic03} which is the most recent interpretation of all
available experimental data. The muonic hydrogen experiment
will help to clarify this undefined situation.

\begin{figure}[h]
\begin{center}
\includegraphics[width=8cm,height=6cm]{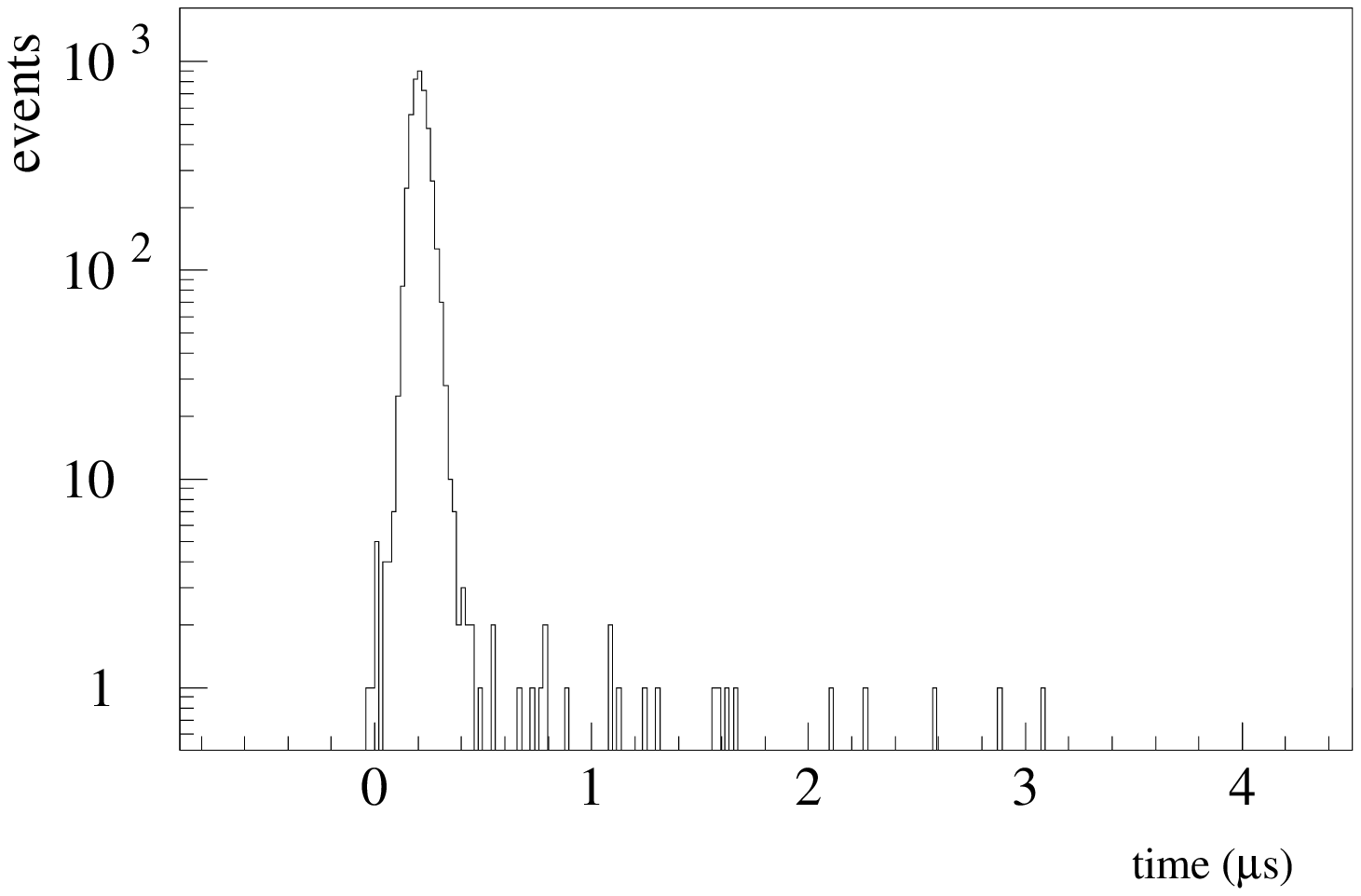}
\end{center}
\vspace*{-.4cm}
\caption{Prompt and delayed x-ray rate in $\mu$p after muon capture. Accum. time: 1.5 h.}
\end{figure}

\noindent
The experiment is performed as follows:
about 60$/$s low energy ($<$ 20 keV) $\mu^-$ are stopped in  low pressure (0.5 mbar) hydrogen gas. After capture in n $\approx$ 14 approximately
1~$\%$ end up as thermalized muonic hydrogen atoms in the meta-stable 2s-state. There they quench via molecular formation with a decay time exceeding 1 $\mu$s. 
About 1.5 $\mu$s after the prompt transitions a 6 $\mu$m mid-infrared laser pulse drives the 2s$\rightarrow$2p transition. The corresponding 2p$\rightarrow$1s
decay x-rays are detected relative to the laser wavelength position. The expected signal rate is 2 times the measured background rate (see Figure 1).
This makes this experiment very promising.

\newabstract 
\label{abs:Juge}

\begin{center}
{\large\bf Lattice QCD calculations of the I=2 pi-pi scattering length}\\[0.5cm]
K.J.~Juge\\[0.3cm]
(BGR-Collaboration)\\[0.3cm]
School of Mathematics, Trinity College, Dublin 2, Ireland
\end{center}

Recent lattice QCD calculations of the I=2 $\pi\pi$ scattering length was 
presented. In particular, the finite volume method of L\"uscher \cite{Luscher}
 was used to extract the $L=0$ scattering length with the parametrized fixed 
point action \cite{spectrum} in the quenched approximation. The highly 
improved action allowed the calculation to be performed on coarse lattices 
with nearly chiral fermions. Signals were obtained down to pion masses of 
roughly $300$ MeV\cite{lat03}. This was the first calculation with chiral 
fermions. Previous calculations used the Wilson action
 \citefour{Sharpe}{Fukugita}{JLQCD}{CPPACS}, improved Wilson action \cite{Liu}
 and/or staggered fermions. Implications on the chiral \cite{Colangelo} and 
quenched chiral behaviour \cite{Bernard} was discussed as well as the recent 
progress made in the $N_f=2$, unquenched calculation \cite{full} with Wilson 
fermions.
\begin{figure}[h]
\begin{center}
\includegraphics[width=69mm]{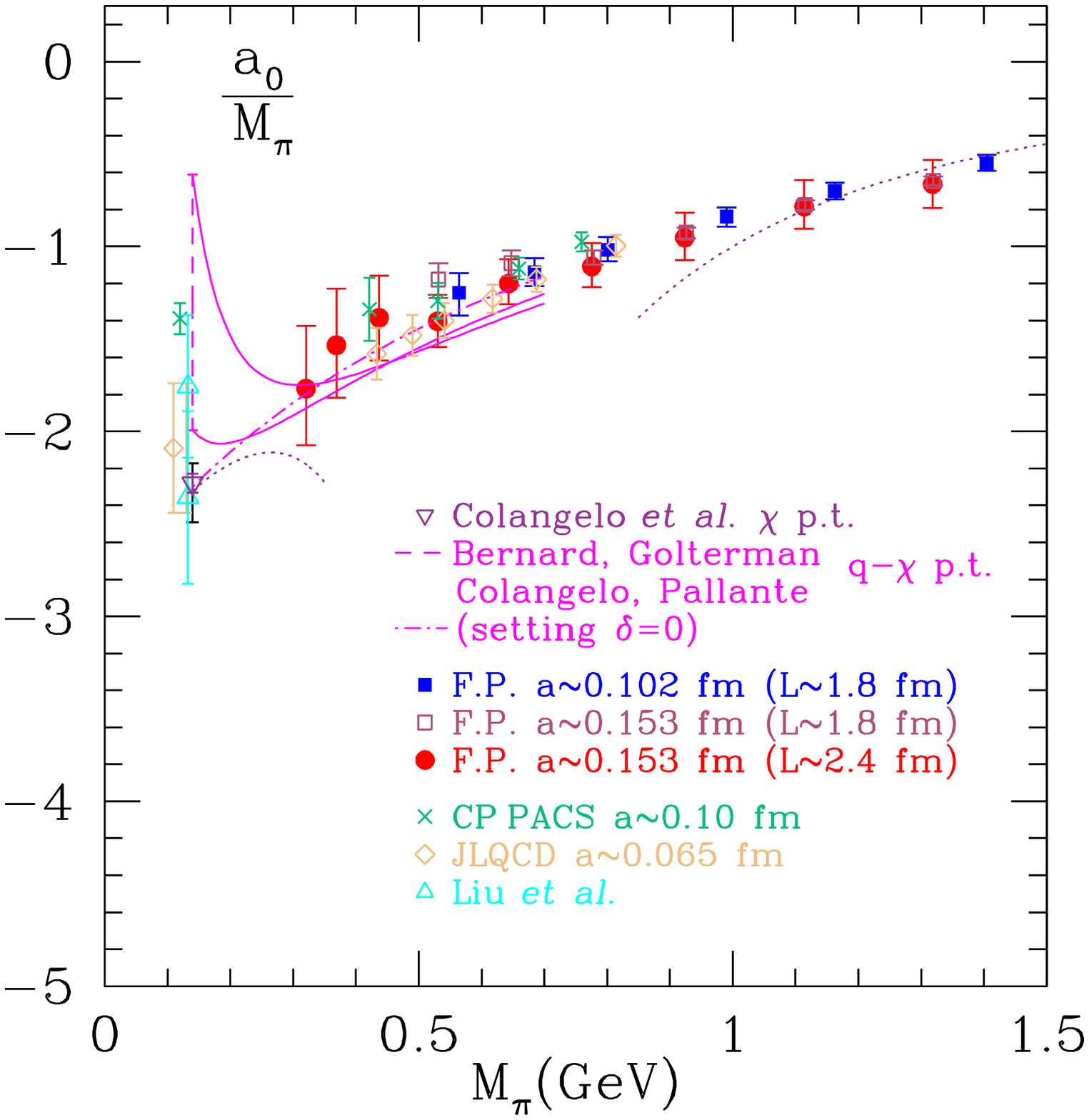}
\end{center}
\label{fig1_juge}
\end{figure}

\newabstract 
\label{abs:Duerr}

\begin{center}
{\large\bf Towards a lattice determination of QCD low-energy constants}\\[0.4cm]
{Stephan D\"urr}$^1$\\[0.0cm]
$^1$DESY Zeuthen, 15738 Zeuthen, Germany\\[0.5cm]
\end{center}

QCD low-energy constants as they appear in Chiral Perturbation Theory
\cite{Gasser:1983yg} fall into two categories: Those which determine how
low-energy Green's functions depend on the external momenta and those which
specify how they depend on the quark masses.
For the latter there is a fair chance that (Euclidean) lattice data will help
to pin them down accurately.

In the past, attempts have been limited by the quenched approximation
\cite{Heitger:2000ay}, but with the current generation of $N_\mathrm{f}\!=\!2$
dynamical data, a new attempt seemed worth while.
Based on recent data by the CP-PACS \cite{AliKhan:2001tx} collaboration, I have
compared the NLO prediction for the degenerate ($m\!=\!m_u\!=\!m_d$) quark mass
dependence of the pseudo-Goldstone boson mass
\cite{Gasser:1983yg}
\vspace*{-1.0mm}
\begin{equation}
\frac{\tilde M_\pi^2}{2\tilde m}=\tilde B
\Big(1+
\frac{\tilde M_\pi^2}{32\pi \tilde F_\pi^2}
\log(\frac{\tilde M_\pi^2}{\tilde\Lambda_3^2})
\Big)
\label{MpiNLO}
\vspace*{-1.0mm}
\end{equation}
to the perturbatively renormalized data (at 1-loop).
In (\ref{MpiNLO}) all energies are in units of the Sommer scale
$r_0^{-1}\!\simeq\!0.4\,\mathrm{GeV}$, i.e.\ $\tilde M_\pi\!=\!M_\pi r_0$.
In addition, I have taken the liberty to combine the chiral log and the
low-energy constant $l_3^\mathrm{r}(\mu)$ into something manifestly
scale-invariant.
The chiral prediction with
$l_3^\mathrm{r}(\mu\!\sim\!0.77\,\mathrm{GeV})\!=\!(0.8\pm3.8)10^{-3}$
(equivalent to $\Lambda_3\!=\!0.6\stackrel{-0.4}{_{+1.4}}\,\mathrm{GeV}$)
taken from \cite{Gasser:1983yg} results in a band which widens rapidly.
In comparison, the theoretical uncertainties (the two renormalized quark masses
should agree) seem less severe.

The main problem is that a phenomenological analysis suggests that -- at
best -- the most chiral lattice points lie within the permissible range for a
NLO chiral fit \cite{Durr}.
If one bluntly ignores this warning and nonetheless attempts a NLO fit to the
entire set, the
\begin{minipage}{9.2cm}
\includegraphics[width=9.0cm]{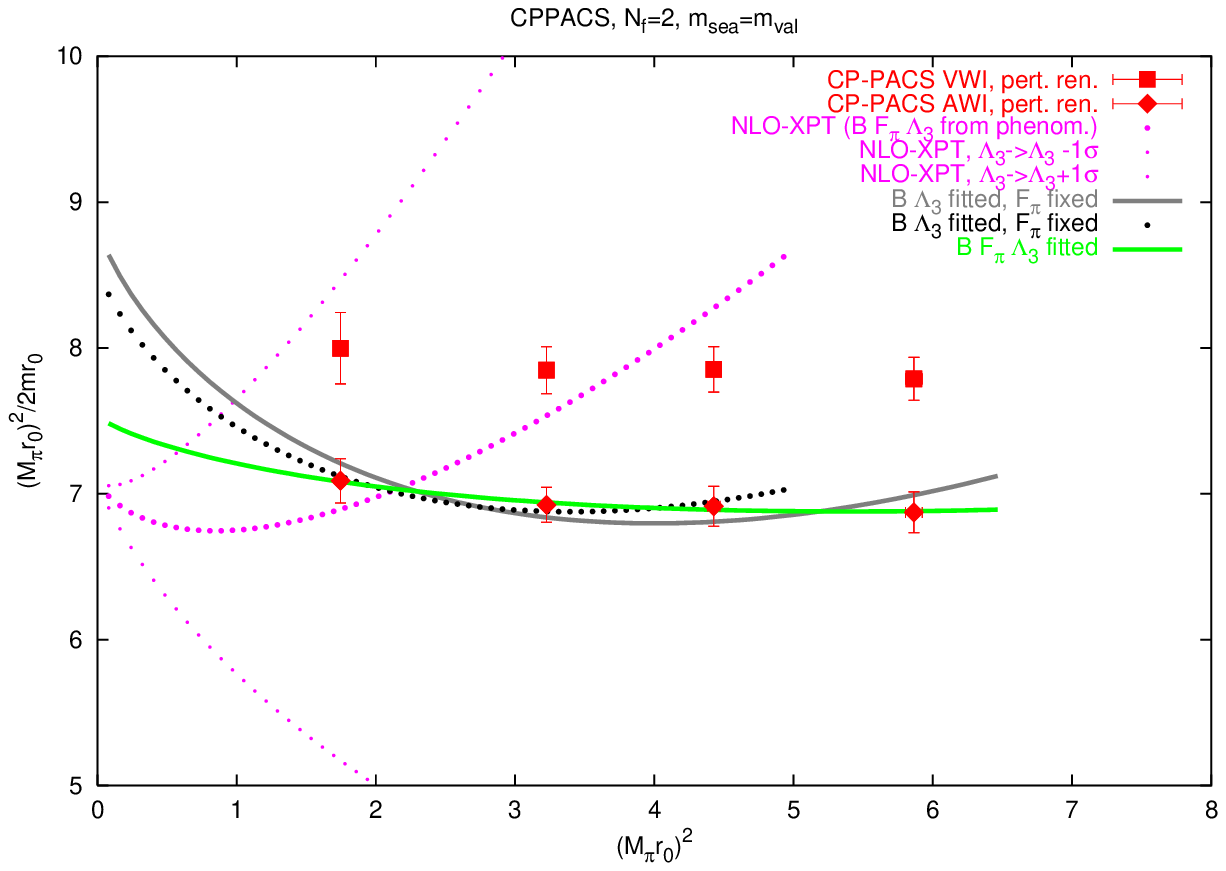}
\end{minipage}
\begin{minipage}{7.2cm}
\vspace*{1mm}
resulting parameters seem reasonable -- except $F_\pi$ which is a bit high
\cite{Durr}.
Keeping $F_\pi\!=\!92.4\,\mathrm{MeV}$ fixed, the fit is still acceptable and
stable under omitting the heaviest point \cite{Durr}.
These findings look inconsistent with statements in \cite{Aoki:2003yv}, but the
main difference is whether $r_0$ (assu\-med independent of $m_\mathrm{sea}$) or
$M_\rho$ (dito) is used to set the scale.
Obviously, the conclusion is that it is too early to use the output of current
fits in phenomenology, but the time will come when $l_3, l_4$ will be reliably
determined on the lattice.
\end{minipage}

\newabstract 
\label{abs:Weise}

\begin{center}
{\large\bf Nucleon mass, sigma term and lattice QCD: chiral extrapolations}\\[0.5cm]
Massimiliano Procura$^{1,2}$, Thomas Hemmert$^2$ and {\bf Wolfram Weise}$^{1,2}$\\[0.3cm]
$^1$ECT*, I-38050 Villazzano (Trento), Italy\\[0.1cm]
$^2$Physik-Department, Technische Universit\"at M\"{u}nchen,
D-85747 Garching, Germany
\end{center}
Lattice QCD on one side and chiral effective field theory, on the other, are progressively developing as important tools to deal with the non-perturbative nature of low-energy QCD and the structure of hadrons. The merger of both strategies has recently been applied to extract
physical properties of hadrons, such as the nucleon, from lattice QCD simulations. Of particular interest in such extrapolations is the detailed quark mass dependence of nucleon properties. Examples are the nucleon mass,  its axial vector coupling constant and magnetic moments \citetwo{AUS}{MUN}.
Accurate computations of the nucleon mass with dynamical fermions and two active flavours are now possible. However, the masses of $u-$ and $d-\,$quarks used in these evaluations exceed their commonly accepted small physical values by typically an order of magnitude. It is at this point where chiral effective field theory methods are useful - within limitations discussed extensively in the literature - in order to interpolate between lattice results, actual observables and the chiral limit ($m_{u,d} \to 0$).
An improved update \cite{proc} is given concerning the quark mass dependence of the nucleon mass and the pion-nucleon sigma term $\sigma_N$. The framework is relativistic $SU(2)_f$ baryon chiral perturbation theory as described in ref.\cite{BL}. The extrapolation to two-flavour lattice QCD results is
performed to order $p^4$. We find $M_0= 885 \pm 6\,{\rm{MeV}} $ for the nucleon mass in the chiral limit ($m_{u,d} \to 0$) and $\sigma_N= 47 \pm 3\,{\rm{MeV}} $  for the sigma term at the physical value of the pion mass (where the errors indicate statistical uncertainties only).

\end{document}